\documentclass{aa}  

\usepackage{graphicx}
\usepackage{txfonts}
\usepackage{natbib}
\usepackage{hyperref}
\usepackage[svgnames]{xcolor}
\usepackage{multirow}
\usepackage{threeparttable}
\usepackage[switch]{lineno}

\bibpunct{(}{)}{;}{a}{}{,}

\begin{document}

   \title{Stellar Population Astrophysics (SPA) with the TNG\thanks{Based on observations made with the Italian Telescopio
Nazionale Galileo (TNG) operated on the island of La Palma by
the Fundaci\'on Galileo Galilei of the INAF (Istituto Nazionale
di Astrofisica) at the Observatorio del Roque de los Muchachos.}}

   \subtitle{NLTE atmospheric parameters and abundances of giant stars in 33 Open Clusters}

\author{M. Dal Ponte
          \inst{1}
          \and
          V. D'Orazi\inst{2,1,3}
          \and
          A. Bragaglia\inst{4}
          \and
          A. R. Casey\inst{5,6,7}
          \and
          N. Storm\inst{8}
          \and
          L. Spina\inst{9,1}
          \and
          J. Alonso-Santiago\inst{10}
          \and         
          G. Andreuzzi\inst{11,12}
          \and
          A. Frasca\inst{10}
          \and
          J. Kos\inst{13}
          \and
          S. Lucatello\inst{1}
          \and
          D. Romano\inst{4}
          \and
          A. Vallenari\inst{1}
          \and
          N. Vernekar\inst{1}
    }

   \institute{INAF - Ossevatorio Astronomico di Padova, vicolo dell’Osservatorio 5, 35122 Padova, Italy \\
    \email{marina.dalponte@inaf.it,valentina.dorazi@inaf.it}
    \and
    Department of Physics, University of Rome Tor Vergata, via della Ricerca Scientifica 1, 00133 Rome, Italy
    \and
    Department of Astronomy \& McDonald Observatory, The University of Texas at Austin, 2515 Speedway, Austin, TX 78712, USA
    \and
    INAF - Osservatorio di Astrofisica e Scienza dello Spazio di Bologna, via P. Gobetti 93/3, 40129 Bologna, Italy 
    \and
    Department of Physics and Astronomy, Monash University, Clayton, VIC 3800, Australia
    \and
    ARC Centre of Excellence for All Sky Astrophysics in 3 Dimensions (ASTRO 3D), Australia
    \and
    Center for Computational Astrophysics, Flatiron Institute, 162 5th Avenue, New York, NY 10010, USA
    \and
    Max Planck Institute for Astronomy,  K\"onigstuhl 17, 69117 Heidelberg, Germany
    \and
    INAF-Osservatorio Astrofisico di Arcetri, Largo E. Fermi 5, 50125, Firenze, Italy
    \and
    INAF - Ossservatorio Astrofisico di Catania, via di S. Sofia 78, I-95123, Catania, Italy
    \and
    INAF - Osservatorio Astronomico di Roma, via Frascati 33, 00178, Monte Porzio Catone, Italy
   \and
   Fundac\'ion Galileo Galilei - INAF, Rambla Jos\'e Ana Fern\'andez P\'erez 7, 38712, Bre\~{n}a Baja, Tenerife, Spain
   \and
   Faculty of Mathematics and Physics, University of Ljubljana, Jadranska 19, 1000 Ljubljana, Slovenia
 }

\date{Received XX; acceptedXX }

 
  \abstract
   {Open clusters serve as important tools for accurately studying the chemical evolution of the Milky Way. By combining precise chemical data from high-resolution spectra with information on their distances and ages, we can effectively uncover the processes that have shaped our Galaxy.}
  {This study aims to derive non-local thermodynamic equilibrium (NLTE) atmospheric parameters and chemical abundances for approximately one hundred giant stars across 33 open clusters with near-solar metallicity. The clusters span a wide range of ages, enabling an assessment of the presence and extent of any age-related abundance gradients.}
   {In the Stellar Population Astrophysics (SPA) project, we acquired new high-resolution spectra of giant stars in a sample of open clusters using the HARPS-N echelle spectrograph at the Telescopio Nazionale Galileo (TNG). We chemically characterized nine open clusters for the first time and reanalyzed previously studied SPA clusters, resulting in a consistent and homogeneous sample.}
   {We determined NLTE atmospheric parameters using the equivalent width method and derived NLTE chemical abundances through spectral synthesis for various elements, including $\alpha$ elements (Mg, Si, and Ti), light odd-Z elements (Na, Al), iron-peak elements (Mn, Co, and Ni), and neutron-capture elements (Sr, Y, and Eu). Our findings are compared with the existing literature, revealing good agreement. We examine the trends of [X/Fe] versus age, confirming previous observations and the enrichment patterns predicted by nucleosynthesis processes. Positive correlations with age are present for $\alpha$ elements such as Mg, Si, Ti, odd-Z Al, and iron-peak elements Mn, Co, Ni, {and Sr}, while Na and {neutron-capture Y and Eu} show a negative trend. This study emphasizes the significance of NLTE corrections and reinforces the utility of open clusters as tracers of Galactic chemical evolution. Furthermore, we provide a benchmark sample of NLTE abundances for upcoming open cluster surveys within large-scale projects such as 4MOST and WEAVE.}
   {}


   \maketitle

\section{Introduction}


Open clusters (OCs) are groups of stars that originate from the same molecular cloud at practically the same time, therefore, they have similar ages, kinematics, and chemical compositions. As a single stellar population, their ages and distances can be estimated 
with great accuracy \citep{Bragaglia2006, Bossini2019, Monteiro2019, Cantat2020}. OCs are ideal tracers of the structure of the Milky Way, and have long been utilized to investigate the Galactic (thin) disk and its chemo-dynamical evolution over time (e.g. \citealt{Janes1979, Sestito2008, Magrini2009, Frinchaboy2013, Netopil2016, Casamiquela2017, Spina2021}). Large spectroscopic surveys such as Gaia-ESO \citep{Gilmore2012, Randich2013, Randich2022}, GALAH \citep{DeSilva2015}, APOGEE \citep{Majewski2017} and LAMOST \citep{Zhao2012}, coupled with the high-quality data from the Gaia mission \citep{Gaia2016, Gaia2023b}, have significantly improved our understanding of the OC population. As a result of the Gaia mission, a large number of OCs were discovered and characterized, while many candidate clusters were expunged from the catalogs (see e.g., \citealt{Cantat2018a, Castro2019, Liu2019, Hao2022, Castro2020, Hunt2023}). In the coming years, WEAVE \citep{Dalton2018, Jin2024} and 4MOST \citep{deJong2019, Lucatello2023} will observe many more clusters at low and intermediate resolution (R $\approx$ 5000 and 20000) and provide new insights on their properties. Despite this, a significant portion of OCs remain unexplored in detail, particularly in terms of their chemical composition, limiting our ability to fully use these objects to probe the Galactic disk’s chemical and dynamical evolution. 

The Stellar Population Astrophysics (SPA) is an ongoing project based on a Large Programme conducted at the Telescopio Nazionale Galileo (TNG) using the HARPS-N and GIANO-B echelle spectrographs for about 70 nights from 2018 to 2021. Within SPA, we acquired high-resolution spectra of stars near the Sun, covering a broad range of ages and properties. To improve our understanding of the Milky Way's star formation and chemical enrichment history, the program focuses on several topics, such as mapping abundance gradients, cosmic spreads, and other irregularities in the abundances and ratios of individual elements within the Galactic disk. We specifically observed a selection of open clusters by either targeting a few stars—mainly giants—in as many clusters as possible, or by focusing on a larger number of stars—primarily main sequence stars—in a smaller set of clusters. The primary goals of these observations were to determine the metallicity and detailed chemical composition across a broad sample of open clusters, taking advantage of their nature as a single stellar population. This means that even a single star can offer valuable insights into the entire cluster, though observing multiple members enhances the precision of key cluster parameters. Additionally, the project investigated how stellar properties relate to evolutionary stages, examining factors such as surface chemical variations caused by diffusion or mixing processes, as well as the evolution of stellar rotation and activity. Several papers have been published as part of the SPA project, analyzing either giant stars or main sequence stars \citet{Frasca2019, Dorazi2020, Casali2020a, Alonso2021, Zhang2021, Zhang2022, Fanelli2022, Vernekar2024}.
    
In this paper, we present a homogeneous analysis of high-resolution spectra of 95 giant stars across 33 open clusters, including nine clusters that are spectroscopically characterized for the first time by our team. Providing data for these previously understudied OCs is crucial for enhancing our understanding of the Galactic disk. We re-analyze giants from \citet{Casali2020a, Zhang2021, Zhang2022}, and \citet{Alonso2021} and include giants from NGC 2632, which \citet{Dorazi2020} studied in the SPA project but focused only on main sequence stars. By ensuring a homogeneous analysis of all spectra, we expand the SPA sample of open clusters. Furthermore, while the previous analyses assumed local thermodynamic equilibrium (LTE) conditions, which may not be suitable for many elements and temperature/surface gravity ranges, we derive atmospheric parameters and abundances using a non-LTE (NLTE) approach.

The structure of the paper is as follows. Section \ref{sec:obs} presents the observations and describes the selected sample of open clusters. In Sect.~\ref{sec:analyses}, we outline the methods employed to determine stellar parameters and abundances. In Sect. \ref{sec:comp_lit} we present a comparison with the literature. Section \ref{sec:gal_trends} investigates the relationships between abundances and age gradients. Finally, Sect.~\ref{sec:conclusions} provides a summary of our findings and conclusions.

    
\section{Observations and the cluster sample}
\label{sec:obs}

The SPA Open Cluster program (SPA-OC) studies main-sequence and giant stars within open clusters. The main-sequence stars in SPA are primarily located in younger, closer clusters, whereas red giants are typically observed in older and more distant clusters (within about 2 kpc from the Sun). In this work, we focus on evolved stars on the red giant branch (RGB) and red clump (RC). The targets have been selected among the high-probability open cluster members based on \citet{Cantat2018a, Cantat2020}, who relied on astrometry and photometry from Gaia Data Release 2 (DR2).  

We deal here with the analysis of the optical HARPS-N spectra with resolution R $\approx$ 115,000 and spectral coverage between 3800-6900 $\AA$ \citep{Cosentino2012}. The analysis and results based on GIANO-B near-infrared spectra are presented in a series of dedicated papers, e.g. \citet{Shilpa2024a, Shilpa2024b, Jian2024}.

The observations discussed in this study were conducted between July 2018 and April 2023. The spectra were automatically reduced with the HARPS-N pipeline, which includes flat-field and bias corrections, wavelength calibration, extraction of 1D spectra, and corrections for barycentric motion. Exposure times varied depending on the star's magnitude and sky conditions, and we combined the multi-exposure spectra before the analysis.

\begin{figure*}
\centering
\includegraphics[width=\linewidth]{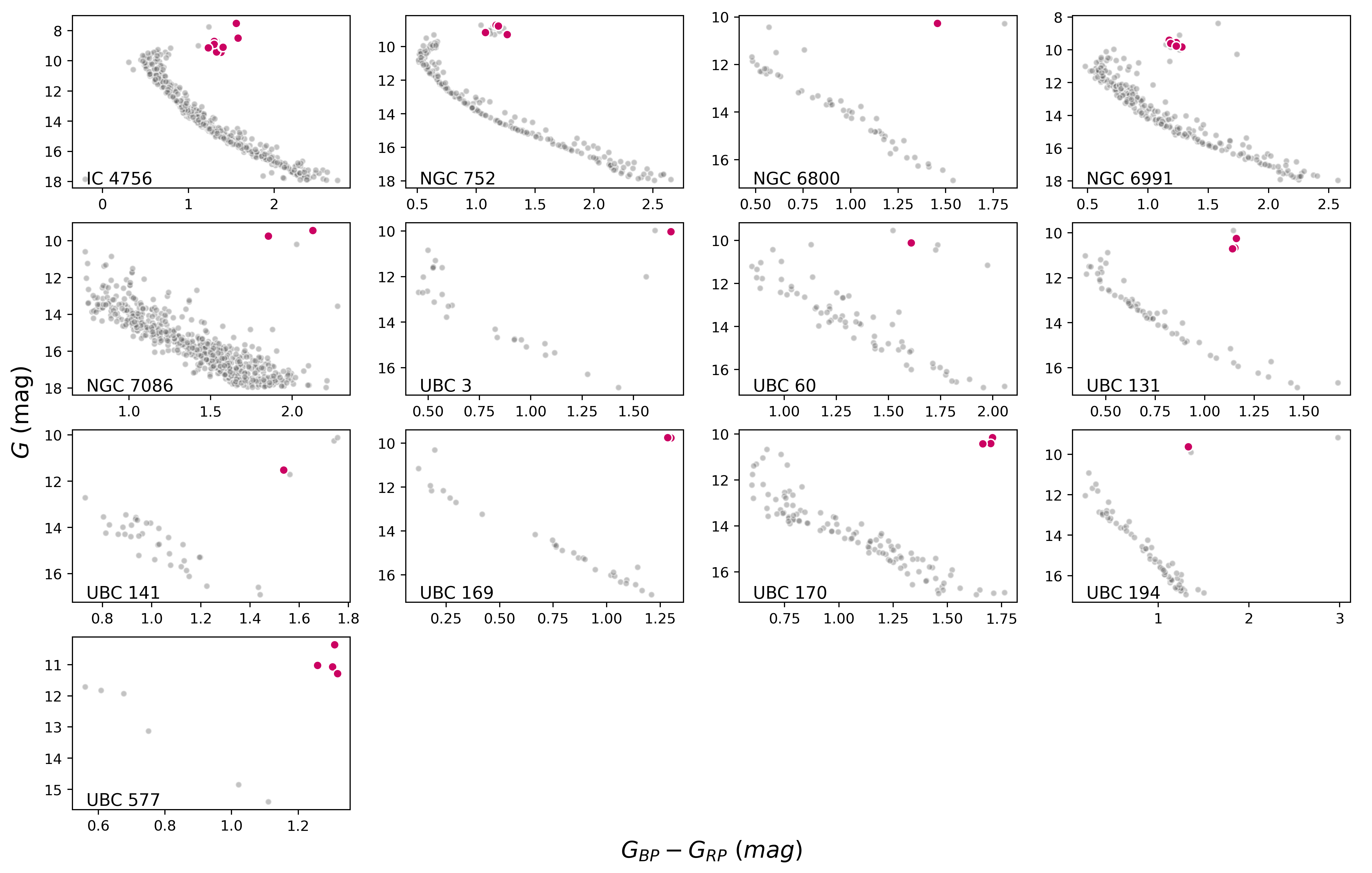}
\caption{Color-magnitude diagrams with Gaia DR3 photometric data ($G$ vs. $G_{BP}-G_{RP}$) for the subset of clusters presented in this study for the first time as part of the SPA project. The red points highlight the member stars observed by SPA.}
\label{fig:cmds}
\end{figure*}

As mentioned before, our sample includes a total of 95 giant stars from 33 open clusters. Table~\ref{tab:lit_sum} displays their names, the number of stars per cluster, and basic parameters such as age, distance, Galactocentric distance, and extinction. The table also indicates references for previous metallicity determinations based on high-resolution spectroscopy; for clusters with high-quality spectroscopy listed in \citet{Netopil2016} we do not provide the references to the original papers. Note that some of these metallicities were not available at the time of observation (see e.g., UBC~3, UBC~141, and UBC~577) and that some metallicities are based on low or intermediate resolution spectroscopy (see below). 

Among the 33 open clusters discussed in this study, 20 were previously analyzed in SPA papers \citep{Casali2020a, Dorazi2020, Zhang2021, Zhang2022, Alonso2021}, see references in Table~\ref{tab:lit_sum} \footnote{The cluster Collinder~350 is analyzed in both \citet{Casali2020a} and \citet{Zhang2021}.}. Four clusters have been reported in the literature by other authors but are presented here for the first time in the context of the SPA project (see Fig.~\ref{fig:cmds} and Table~\ref{tab:lit_sum}). Finally, this is the first metallicity determination based on high-resolution spectroscopy for nine OCs (NGC~6800, UBC~131, UBC~169, UBC~170, UBC~194 without metallicity values and NGC~7086, UBC~60, UBC~141, UBC~577 with values based on low-resolution LAMOST or medium-resolution Gaia-RVS spectra).

The inclusion of the 20 open clusters already studied in the SPA framework allows us to have a broader sample analyzed homogeneously and to extend the age range. These 33 open clusters in the SPA sample are generally located close to the Galactic midplane and to the Sun (average Galactocentric distance of 8.6 kpc, with $R_{\textrm{GC}}$ spanning from 7 kpc to approximately 10 kpc) and have ages ranging from about 40 Myr to 4 Gyr (see Table~\ref{tab:lit_sum}). 

Figure \ref{fig:cmds} presents the color-magnitude diagrams using Gaia DR3 photometric data for the 13 open clusters that have not been included in previous SPA analyses. We also provide a brief description of these clusters in the following paragraphs. 

\begin{table*}
\centering 
\caption{Summary of the clusters analyzed in this work, including their properties and the number of stars in each cluster (N). Age, A$_V$, distance, and $R_{\textrm{GC}}$ are from \citet{Cantat2020}.}
\label{tab:lit_sum}
\scalebox{0.95}{
\begin{tabular}{llcccrrl}
\hline\hline
\multicolumn{1}{c}{Cluster} & \multicolumn{1}{c}{Other names} & N & \multicolumn{1}{c}{Age (Gyr)} & \multicolumn{1}{c}{A$_V$ (mag)} & \multicolumn{1}{c}{Dist. (pc)} & \multicolumn{1}{c}{$R_{\textrm{GC}}$ (pc)} & \multicolumn{1}{c}{[Fe/H] References} \\
\hline
Alessi 1 & Casado-Alessi 1 & 4 & 1.45 & 0.08 & 689 & 8726 & 1,2 \\		   
Alessi-Teutsch 11 & ASCC 112, Alessi 46 & 1 & 0.14 & 0.37 & 634 & 8335 & 1 \\
Basel 11b & FSR 877 & 3 & 0.23 & 1.56 & 1793 & 10121 & 1,3 \\
COIN-Gaia 30 & - & 1 & 0.26 & 1.25 & 767 & 8804 & 1	\\
Collinder 350 &	- & 2 & 0.59 & 0.52 & 371 & 8021 & 1,4 \\
Collinder 463 &	- & 2 & 0.11 & 0.79 & 849 & 8874 & 1 \\
Gulliver 18	& Collinder 416	& 1 & 0.04 & 1.59 & 1595 & 7816 & 1	\\
Gulliver 24	& - & 1 & 0.18 & 1.05 & 1498 & 9131 & 1	\\
Gulliver 51	& - & 1 & 0.36 & 1.42 & 1536 & 9410 & 4 \\
IC 4756 & Collinder386, Melotte 210 & 12 & 1.29 & 0.29 &  506 & 7938 & 2,5,6,7,8,9 \\
NGC 2437 & M 46, Melotte75 & 5 & 0.30 & 0.73 & 1511 & 9345 & 1,7 \\
NGC 2509 & Melotte 81, Collinder 171 & 1 & 1.51 & 0.23 & 2495 & 9887 & 1 \\
NGC 2548 & M 48, Melotte 85 & 3 & 0.40 & 0.15 & 772 & 8857 & 1,7,9,10,11 \\
NGC 2632 & M 44, Praesepe & 2 & 0.68 & 0.00 & 183 & 8479 & 2,3,5,9,10,12,13,14 \\
NGC 2682 & M 67 & 2 & 4.27 & 0.07 &  889 & 8964 & 1,2,3,5,7,9 \\
NGC 6800 & - & 1 & 0.41 & 0.83 & 1012 & 7868 & - \\
NGC 6991 & - & 5 & 1.55 & 0.20 & 577 & 8333 & 9,15 \\
NGC 7044 & Collinder 433 & 4 & 1.66 & 1.78 & 3252 & 8729 & 2,4,16 \\
NGC 7086 & Collinder 437 & 2 & 0.19 & 1.81 & 1677 & 8632 & 17 \\
NGC 7209 & Melotte 238, Collinder 444 & 2 & 0.43 & 0.53 & 1154 & 8525 & 1 \\
NGC 752 & Melotte 12, Theia 1214 & 5 & 1.17 & 0.07 & 483 & 8669 & 3,5,9,18,19 \\
Ruprecht 171 & - & 7 & 2.75 & 0.68 & 1522 & 6895 & 4,9,16 \\
Stock 2 & - & 8 & 0.40 & 0.50 & 399 & 8619 & 15,20 \\
Tombaugh 5 & - & 3 & 0.19 & 2.07 & 1706 & 9768 & 1,21 \\
UBC 3 & Alessi 161 & 1 & 0.13 & 0.98 & 1704 & 7223 & 9 \\
UBC 60 & COIN-Gaia 11 & 1 & 0.79 & 1.25 & 669 & 8977 & 17 \\
UBC 131 & Alessi 116, UPK 84 & 3 & 1.00 & 0.31 & 944 & 7980 & 2	\\
UBC 141 & - & 1 & 2.09 & 0.51 & 1315 & 8176 & 2,17 \\
UBC 169 & - & 2 & 0.30 & 0.26 & 1347 & 8749 & 17 \\
UBC 170 & FoF 1800, LP 1800	& 3 & 0.33 & 1.25 & 1392 & 8809 & - \\
UBC 194 & - & 1 & 0.23 & 0.48 & 1361 & 9394 & - \\
UBC 577 & Alessi 191 & 4 & 2.75 & 0.00 & 1122 & 7718 & 2 \\
UPK 219 & - & 1 & 0.15 & 1.20 & 873 & 8735 & 1 \\
\hline
\end{tabular}
}
\tablefoot{References: (1) \citet{Zhang2021}; (2) \citet{Viscasillas2023}; (3) \citet{Myers2022} ; (4) \citet{Casali2020a};
(5) \citet{Netopil2016}, High-quality spectroscopic metallicity, see original references there; (6) \citet{Bagdonas2018}; 
(7) \citet{Ray2022}; (8) \citet{Tsantaki2023}; (9) \citet{CarbajoHijarrubia2024}; (10) \citet{Spina2021}; (11) \citet{Ramos2024};
(12) \citet{Dorazi2020}; (13) \citet{Cummings2017}; (14) \citet{Vejar2021}; (15) \citet{Reddy2019}; (16) \citet{Seshashayana2024};
(17) \citet{Fu2022}, LAMOST low-resolution spectra; (18) \citet{Lum2019}; (19) \citet{Bocek2015}; (20) \citet{Alonso2021}; (21) \citet{Baratella2018}.  }
\end{table*}

\noindent \textbf{IC 4756 -} This intermediate-age cluster (about 1.3 Gyr old, according to \citealt{Cantat2020}, 970 Myr according to \citealt{Bossini2019}) is located very close to the Sun ($\approx$ 500 pc) at a Galactocentric distance of $\approx$ 8 kpc. Several studies have assessed its metallicity, with a few of them also investigating its detailed elemental abundances \citep{Jacobson2007, Santos2009, Smiljanic2009, Pace2010, Ting2012, Bagdonas2018, Casamiquela2017, Casamiquela2019, Ray2022, Tsantaki2023, CarbajoHijarrubia2024}. We have eight stars in common with the existing literature, which will later be used for comparison with our results (details can be found in Sect.~\ref{sec:comp_lit}). \\

\noindent \textbf{NGC 752 (and Theia 1214) -} This cluster is located near the Sun and has an intermediate age—approximately 1.2 Gyr (\citealt{Cantat2020}) or 1.48 Gyr (\citealt{Bossini2019}). It has been extensively studied, particularly for lithium measurements, as discussed in works such as \citet{Boesgaard2022} and related references. Other metallicity and abundance studies include \citet{Sestito2004, Carrera2011, Reddy2012, Bocek2015, Casamiquela2017, Casamiquela2019, Lum2019, Spina2021, Myers2022, Donor2020, CarbajoHijarrubia2024}. The cluster shows signs of either dynamical or tidal dispersal \citep{Carraro2014} and features long, asymmetric tidal tails extending over 260 pc on the sky \citep{Bhattacharya2021, Boffin2022, Kos2024}. In addition, Theia~1214, one of the strings discovered by \citet{Kounkel2019}, is likely associated with NGC 752. We added to the sample two stars that are part of Theia~1214 according to \citet{Kounkel2019}. However, in this work, we consider them as members of NGC 752. In total, we have five stars, where the three stars observed in the main body of the cluster are in common with the literature.\\

\noindent \textbf{NGC 6800 -} Located at approximately 8 kpc and with an age of around 400 Myr (\citealt{Cantat2020} or 345 Myr, \citealt{Bossini2019}), this OC has not yet been chemically characterized in the literature. \\

\noindent \textbf{NGC 6991 -} It has an intermediate age of roughly 1.5 Gyr and is located about 580 pc from the Sun at a Galactocentric distance of approximately 8.4 kpc. Its metallicity has been derived by \citet{Reddy2019} and by the OCCASO project \citep{Casamiquela2017, Casamiquela2019, CarbajoHijarrubia2024}. We have only one star in common for comparison. \\

\noindent \textbf{NGC 7086 -} Located at a Galactocentric distance of approximately 8.6 kpc, this young OC ($\approx$ 200 Myr old, \citealt{Cantat2020}) has not yet been chemically characterized in the literature. The only reference we found is in \citet{Fu2022},  based on the LAMOST low-resolution spectra. \\

\noindent \textbf{The ``new" Gaia clusters -} UBC~3, UBC~60/COIN-Gaia~11,  UBC~141, UBC 169, UBC~170, UBC~194, UBC~577, UBC 131/UPK~84, and UPK~219 (see Table~\ref{tab:lit_sum} for alternative names) are new open clusters\footnote{Actually, a few of the UBC and UPK clusters are rediscovery of objects already known, see the cited studies for the cross-matching with older catalogs.} discovered by \citet{Castro2018, Castro2019, Castro2020} (the UBC clusters) and \citet{Sim2019} (the UPK clusters) using Gaia DR2 and DR3 data. For the UBC clusters, a clustering algorithm (DBSCAN) was used to detect overdensities in the astrometric space (position, parallax, and proper motions), followed by a neural network classifier to separate random clusterings from true clusters. For the UPK clusters, \citet{Sim2019} relied on a visual inspection of proper motion and spatial distributions. Most of these clusters were not yet studied spectroscopically, although UBC~60, UBC~131 UBC~141, and UBC~169 have metallicity based on LAMOST low-resolution spectra \citep{Fu2022} or on intermediate-resolution Gaia-RVS spectra \citet{Viscasillas2023}. Only  UBC~3 has been fully chemically characterized within the high-resolution project OCCASO \citep{CarbajoHijarrubia2024}. 

\begin{table*}
\centering
\caption{Stellar parameters obtained with {\tt LOTUS} for the 95 stars presented in this work.}
\label{tab:atm_params}
\begin{tabular}{llccrc}
\hline \hline
\multicolumn{1}{c}{Star} & \multicolumn{1}{c}{Gaia DR3 ID} & \multicolumn{1}{c}{$T_{\rm eff}$ (K)} & \multicolumn{1}{c}{log $g$ (dex)} & \multicolumn{1}{c}{[Fe/H] (dex)} & \multicolumn{1}{c}{$\xi$ (km$\rm s^{-1}$)} \\
\hline
Alessi 1 \#2 & 402506369136008832 & 4986 $\pm$ 30 & 2.83 $\pm$ 0.07 & -0.00 $\pm$ 0.01 & 1.30 $\pm$ 0.01 \\
Alessi 1 \#3 & 402505991180022528 & 4996 $\pm$ 30 & 2.84 $\pm$ 0.07 & -0.01 $\pm$ 0.01 & 1.28 $\pm$ 0.01 \\
Alessi 1 \#5 & 402867593065772288 & 4939 $\pm$ 31 & 2.73 $\pm$ 0.08 & -0.03 $\pm$ 0.01 & 1.35 $\pm$ 0.01 \\
Alessi 1 \#6 & 402880684126058880 & 4985 $\pm$ 32 & 2.87 $\pm$ 0.08 & -0.01 $\pm$ 0.01 & 1.28 $\pm$ 0.01 \\
Alessi Teutsch 11 \#1 & 2184332753719499904 & 4517 $\pm$ 35 & 2.15 $\pm$ 0.11 & -0.04 $\pm$ 0.01 & 1.76 $\pm$ 0.01 \\
Basel 11b \#1 & 3424056131485038592 & 4950 $\pm$ 35 & 2.51 $\pm$ 0.09 & 0.01 $\pm$ 0.01 & 1.81 $\pm$ 0.01 \\
Basel 11b \#2 & 3424055921028900736 & 5008 $\pm$ 33 & 2.64 $\pm$ 0.08 & 0.02 $\pm$ 0.01 & 1.77 $\pm$ 0.01 \\
Basel 11b \#3 & 3424057540234289408 & 4760 $\pm$ 34 & 2.55 $\pm$ 0.10 & 0.08 $\pm$ 0.01 & 1.47 $\pm$ 0.02 \\
COIN-Gaia 30 \#1 & 532533682228608384 & 4796 $\pm$ 35 & 2.41 $\pm$ 0.10 & -0.04 $\pm$ 0.01 & 1.69 $\pm$ 0.01 \\
Collinder 350 \#1 & 4372743213795720704 & 4197 $\pm$ 53 & 1.65 $\pm$ 0.19 & -0.15 $\pm$ 0.02 & 1.86 $\pm$ 0.02 \\
\hline
\end{tabular}
\tablefoot{Only an excerpt of the table is shown here. The full table is available at the CDS.}
\end{table*}

\section{Spectroscopic analysis}
\label{sec:analyses}

We first corrected the spectra for radial velocity (RV) and performed continuum normalization. For this purpose, we used the {\tt iSpec} tool \citep{BlancoCuaresma2014}. The continuum normalization was carried out using a spline function with 60 splines of quadratic order. Table \ref{tab:sample_prop} provides an overview of the properties of our giant stars in 14 open clusters. We refer to \citet{Casali2020a, Zhang2021, Zhang2022, Alonso2021} for the remainder of the sample. This table includes, for each star, Gaia DR3 \citep{Gaia2023a} right ascension and declination, magnitudes, and RV, along with the total exposure times, the resulting signal-to-noise ratio (S/N), and the estimated RV and associated error obtained from {\tt iSpec}. There is a remarkable agreement between the RV measurements from Gaia DR3 and those from {\tt iSpec}, with a mean difference of +0.08 km~s$^{-1}$. The comparison was done after the exclusion of a few binary stars, either already known or found by us based on RV differences (see Sect.~\ref{sec:comp_lit} for more details).

\subsection{Stellar parameters}
\label{sec:stellar_param}

Spectroscopic parameters ($T_{\rm eff}$, $\log g$, metallicity [Fe/H]\footnote{We adopt the standard spectroscopic notation of [Fe/H]=$\log \frac{N(Fe)_\ast}{N(H)_\ast} - \log \frac{N(Fe)_\odot}{N(H)_\odot}$} and microturbulent velocity $\xi$) were obtained using the equivalent width (EW) method. The line list for Fe~{\sc i} and {\sc ii} transitions used in this work is based on \citet{Li2023}. These authors combine iron lines from the Gaia-ESO line list \citep{Jofre2014, Heiter2015, Heiter2021} and from the R-Process Alliance (RPA) survey \citep{Hansen2018, Sakari2018, Ezzeddine2020, Holmbeck2020}. In total, we have 165 Fe~{\sc i} \ and 20 Fe ~{\sc ii} \ lines.
The EWs were measured using the {\tt smhr} tool \footnote{\tt \hyperlink{https://github.com/andycasey/smhr}{https://github.com/andycasey/smhr} } \citep{Casey2014}. Our method involved measuring the EW for a single reference star (\#5 from IC 4756) using {\tt smhr}. Subsequently, we calculated the EWs for the rest of the sample using the reference star input masks with the {\tt stellardiff} tool \footnote{\tt \hyperlink{https://github.com/andycasey/stellardiff}{https://github.com/andycasey/stellardiff}}. This code is specifically designed for differential analyses, allowing users to make assumptions about the local continuum around the spectral lines. This functionality helps minimize the effects of poor spectral normalization or unresolved features in the continuum. Additionally, we excluded any line with an EW greater than 200 m\AA~ and with an error exceeding 10 m\AA.

Initially, we employed the {\tt qoyllur-quipu (q2)} LTE differential analysis code \citep{Ramirez2014} on a test subsample of 14 open cluster stars, as routinely performed in previous works such as \cite{Casamiquela2017, Casamiquela2019, liu2016}. The {\tt q2} tool is a Python package that enables users to use {\tt MOOG} \citep{sneden1973} to calculate atmospheric parameters and abundances based on the standard iron line excitation and ionization equilibrium techniques. The code works iteratively to adjust the parameters, aiming to minimize correlations with excitation potential ($\chi$) and the reduced equivalent width (REW), while also reducing the differences between the mean iron abundances derived from Fe~{\sc i} and Fe {\sc ii} lines. A notable strength of this code is its capability to perform line-by-line differential analysis, which offers a more robust method for analyzing ``twin'' stars (as discussed in, for example, \citealt{melendez2009}). For the {\tt q2} method, we initially calculated differential parameters using the IC 4756 \#5 as the reference for the entire sample. However, this approach proved inadequate due to the wide range of effective temperatures ($T_{\rm eff}$) and surface gravities ($\log g$) exhibited by our cluster giants. To address this, we established criteria for selecting multiple reference stars, ensuring that the differences between each star and its chosen reference satisfied the conditions of $\Delta T_{\rm eff} < 150$ K and $\Delta \log g < 0.25$. Ultimately, we identified four suitable reference stars (IC 4756 \#4, \#5, \#13 and UBC 170 \#1). To choose these reference stars, we first ran the {\tt q2} method to compute the absolute parameters for all the stars in the sample. We then used the absolute parameters of the reference stars to apply the {\tt q2} method again to derive the parameters for each subsample with their respective reference stars. However, recognizing the limitations of this method, which still resulted in a significant internal spread within each cluster, exceeding $\approx$0.20 dex (peak-to-valley) for giant stars, we opted to adopt an NLTE methodology and proceeded as follows. 

\begin{figure*}
\centering
\includegraphics[width=\linewidth]{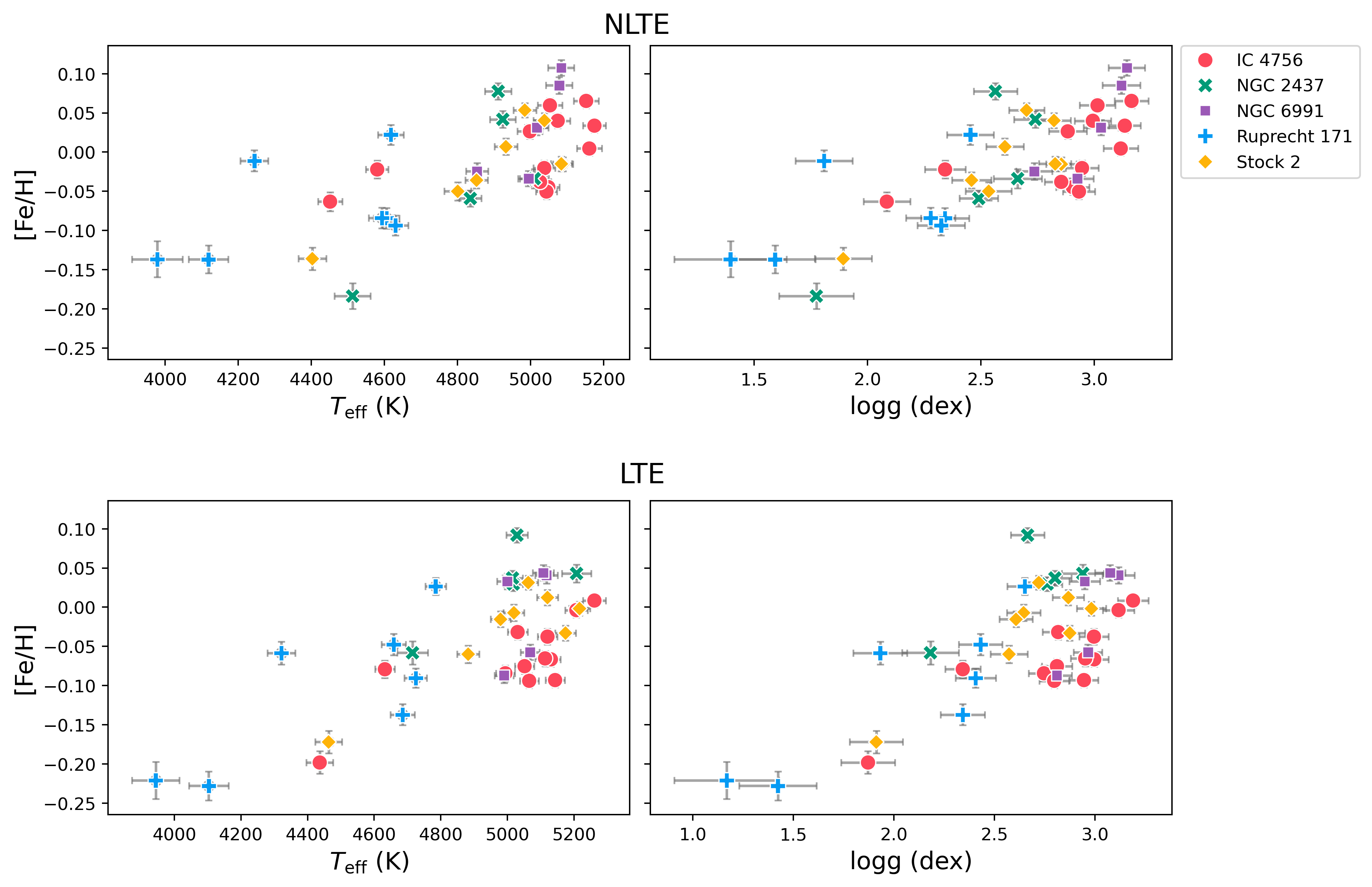}
\caption{Atmospheric parameters [Fe/H], $T_{\rm eff}$, and $\log g$ obtained with {\tt LOTUS} for clusters containing more than five stars. The first row presents the NLTE results, while the second row shows the LTE results.}
\label{fig:lotus}
\end{figure*}

We employed the code {\tt LOTUS} (non-LTE Optimization Tool Utilized for the derivation of atmospheric Stellar parameters, \citealt{Li2023})\footnote{\tt \hyperlink{https://github.com/Li-Yangyang/LOTUS}{https://github.com/Li-Yangyang/LOTUS} }, which allows the derivation of atmospheric parameters ($T_{\rm eff}$, $\log g$, [Fe/H] and $\xi$) based on EW measurements of Fe~{\sc i} and Fe {\sc ii} lines. {\tt LOTUS} performs interpolations of a generalized curve of growth (GCOG) from a grid of theoretical EWs in both LTE and NLTE, following the methodology outlined by \citet{Boeche2016}. Stellar parameters are derived by minimizing the slopes of excitation and ionization equilibrium through an iterative global minimization process, with uncertainties in the atmospheric parameters estimated using a Markov-Chain Monte Carlo algorithm, further details are provided by \citealt{Li2023} (see their Sect. 2.5). Figure \ref{fig:mcmc_lotus} show the corner plot from the posterior distribution of the stellar parameters and their uncertainty for IC 4756 \#5. Users can select either LTE or NLTE modes. When considering cool giants, adopting the NLTE approach over LTE-based analyses is crucial for accurately modeling their atmospheres. These stars have low surface gravities and densities, which prevent them from maintaining collisional equilibrium in their outer layers, causing a departure from LTE. NLTE effects arise from these imbalances, particularly in the ionization balance of elements like iron, and can significantly impact the derived abundances and stellar parameters \citep[see e.g.,][and references therein]{bergemann2012, lind2022}. Therefore, in this study, we chose to run {\tt LOTUS} with NLTE corrections, using the following input parameters to represent the stars in our sample: stellar type K ($T_{\rm eff}$ ranging from 4000 K to 5200 K), giants ($\log g$ between 0.5 dex and 3.0 dex), and metal-rich ([Fe/H] from $-$0.5 to 0.5). We ran {\tt LOTUS} on the 33 SPA open clusters, which comprises 95 giant stars. In Table \ref{tab:atm_params}, we report the derived stellar parameters $T_{\rm eff}$, $\log g$, [Fe/H], and $\xi$ for a subset of the sample. The full table is available at CDS \footnote{\tt \hyperlink{http://cdsweb.u-strasbg.fr/cgi-bin/qcat?J/A+A/}{http://cdsweb.u-strasbg.fr/cgi-bin/qcat?J/A+A/} }. The uncertainties in [Fe/H] were calculated using $\sigma/\sqrt{n}$, where $\sigma$ is the standard deviation of the Fe {\sc i} lines given by {\tt LOTUS}, and $n$ is the total number of Fe {\sc i} lines used in the analysis. 

Figure \ref{fig:lotus} presents the stellar parameters [Fe/H], $T_{\rm eff}$, and $\log g$ for clusters containing more than five giant stars. Looking at stars in the same cluster, a noticeable trend emerges between metallicity and effective temperature, with cooler stars (those with $T_{\rm eff} \approx 4600$ K) generally exhibiting lower metallicities. This trend is also present when considering surface gravity, which is to be expected as $T_{\rm eff}$ and gravity are correlated in the sample. Indeed, were the trend with $T_{\rm eff}$ be removed, the correlation with $\log g$ would disappear as well. This pattern is not new and has been discussed in the literature (see, e.g., \citealt{Casali2020a, Beeson2024}). 

Determining the underlying cause of these trends is complex and may be attributed to several factors, including continuum displacement in cooler giants, the influence of blended features, the selection of line lists and atomic parameters. Recently, \cite{kos2025} reported abundance trends related to stellar parameters, particularly $T_{\rm eff}$, for open clusters in GALAH DR4 \citep{buder2024}. We refer to \cite{kos2025} for a comprehensive discussion of the potential explanations for this trend. Additionally, as also emphasized in \cite{kos2025}, atomic diffusion cannot explain the observed trend, as it operates at different parameters and evolutionary stages, leading to trends that differ from those observed. Finally, a NLTE analysis reduces the discrepancies between warmer and cooler stars within the same clusters; however, it does not completely resolve the issue. This is clearly illustrated in Fig.~\ref{fig:lotus}, where we compare LTE (lower panel) and NLTE parameters obtained using {\texttt LOTUS}. Given the amplitude trends calculated in \cite{kos2025}, our results are consistent with GALAH stars for most elements. However, we observe larger internal variation for Sr and Y, which is primarily driven by a single cluster, Ruprecht 171, where a few cold stars contribute to the discrepancy. 

In cases with a large number of stars per cluster, one might consider detrending these results, as done in \cite{kos2025}. However, given the statistic limitation in our case, we have chosen to present our abundance measurements as they are, recognizing that this trend may introduce an internal precision limitation of less than $\approx$ 0.1 dex.

A direct, full comparison of LTE and NLTE results is not the focus of the present paper and we conclude the exercise by showing the differences between LTE and NLTE parameters in Fig.~\ref{fig:lte_nlte_params}. There, a clear positive trend is present, indicating that stars with larger (and positive) LTE–NLTE temperature differences also show larger differences in surface gravity. This suggests that LTE analysis tends to overestimate both $T_{\rm eff}$ and $\log g$, and that NLTE analysis mitigates these biases, yielding more consistent parameters across different stellar types. 

\begin{table*}
\centering
\caption{NLTE chemical abundances derived with {\tt TSFitPy} for the 95 stars presented in this work.}
\label{tab:abundances}
\setstretch{1.12}
\scalebox{0.63}{
\begin{tabular}{lrrrrrrrrrrr}
\hline \hline
\multicolumn{1}{c}{Star} & \multicolumn{1}{c}{[Mg/Fe]} & \multicolumn{1}{c}{[Si/Fe]} & \multicolumn{1}{c}{[Ti/Fe]} & \multicolumn{1}{c}{[Na/Fe]} & \multicolumn{1}{c}{[Al/Fe]} & \multicolumn{1}{c}{[Mn/Fe]} & \multicolumn{1}{c}{[Co/Fe]} & \multicolumn{1}{c}{[Ni/Fe]} & \multicolumn{1}{c}{[Sr/Fe]} & \multicolumn{1}{c}{[Y/Fe]} & \multicolumn{1}{c}{[Eu/Fe]} \\
\hline
Alessi 1 \#2 & -0.036 $\pm$ 0.05 &  0.040 $\pm$ 0.05 &  0.047 $\pm$ 0.03 &  0.115 $\pm$ 0.01 & -0.042 $\pm$ 0.01 & 0.027 $\pm$ 0.06 & -0.058 $\pm$ 0.07 &  0.067 $\pm$ 0.06 &  0.014 $\pm$ 0.05 & 0.213 $\pm$ 0.03 & - \\
Alessi 1 \#3 & -0.010 $\pm$ 0.05 &  0.012 $\pm$ 0.04 &  0.079 $\pm$ 0.03 &  0.123 $\pm$ 0.01 &  0.004 $\pm$ 0.01 & 0.008 $\pm$ 0.07 & -0.064 $\pm$ 0.06 &  0.078 $\pm$ 0.06 & -0.009 $\pm$ 0.05 & 0.172 $\pm$ 0.02 & 0.143 $\pm$ 0.05 \\
Alessi 1 \#5 & -0.028 $\pm$ 0.05 & -0.045 $\pm$ 0.02 &  0.044 $\pm$ 0.02 &  0.112 $\pm$ 0.01 & -0.024 $\pm$ 0.05 & 0.015 $\pm$ 0.06 & -0.066 $\pm$ 0.06 &  0.045 $\pm$ 0.05 &  0.044 $\pm$ 0.05 & 0.170 $\pm$ 0.03 & - \\
Alessi 1 \#6 & -0.025 $\pm$ 0.05 &  0.015 $\pm$ 0.05 &  0.054 $\pm$ 0.03 &  0.079 $\pm$ 0.01 & -0.045 $\pm$ 0.01 & 0.015 $\pm$ 0.06 & -0.055 $\pm$ 0.06 &  0.092 $\pm$ 0.06 &  0.031 $\pm$ 0.05 & 0.184 $\pm$ 0.03 & 0.05 $\pm$ 0.05\\
Alessi Teutsch 11 \#1 & -0.097 $\pm$ 0.05 &  0.014 $\pm$ 0.02 &  0.024 $\pm$ 0.02 & -0.065 $\pm$ 0.01 & -0.047 $\pm$ 0.02 & -0.086 $\pm$ 0.11 & -0.077 $\pm$ 0.06 & -0.121 $\pm$ 0.03 & -0.041 $\pm$ 0.05 & 0.166 $\pm$ 0.05 & - \\
Basel 11b \#1 & -0.095 $\pm$ 0.05 & -0.091 $\pm$ 0.02 & -0.025 $\pm$ 0.03 &  0.106 $\pm$ 0.01 &  0.003 $\pm$ 0.05 & -0.098 $\pm$ 0.10 & -0.126 $\pm$ 0.07 & -0.057 $\pm$ 0.05 & -0.025 $\pm$ 0.05 & 0.117 $\pm$ 0.05 & 0.153 $\pm$ 0.05 \\
Basel 11b \#2 & -0.064 $\pm$ 0.05 & -0.066 $\pm$ 0.05 & 0.028 $\pm$ 0.02 &  0.150 $\pm$ 0.05 & -0.009 $\pm$ 0.05 & -0.047 $\pm$ 0.06 & -0.083 $\pm$ 0.06 &  0.020 $\pm$ 0.06 &  0.064 $\pm$ 0.05 & 0.239 $\pm$ 0.05 & 0.223 $\pm$ 0.05\\
Basel 11b \#3 & -0.075 $\pm$ 0.05 & -0.085 $\pm$ 0.02 & 0.032 $\pm$ 0.02 & -0.051 $\pm$ 0.02 & -0.039 $\pm$ 0.05 & - & -0.060 $\pm$ 0.06 & -0.027 $\pm$ 0.05 &  0.287 $\pm$ 0.05 & 0.336 $\pm$ 0.06 & 0.121 $\pm$ 0.05 \\
COIN-Gaia 30 \#1 & -0.067 $\pm$ 0.05 & -0.093 $\pm$ 0.02 &  0.012 $\pm$ 0.03 &  0.093 $\pm$ 0.01 & -0.047 $\pm$ 0.05 & -0.115 $\pm$ 0.09 & -0.098 $\pm$ 0.06 & -0.080 $\pm$ 0.05 &  0.119 $\pm$ 0.05 & 0.205 $\pm$ 0.05 & 0.132 $\pm$ 0.05 \\
Collinder 350 \#1& -0.034 $\pm$ 0.05 &  0.019 $\pm$ 0.03 &  0.063 $\pm$ 0.03 &  0.146 $\pm$ 0.04 &  0.020 $\pm$ 0.03 & -0.201 $\pm$ 0.13 & -0.032 $\pm$ 0.06 & -0.201 $\pm$ 0.13 &  0.303 $\pm$ 0.05 & 0.440 $\pm$ 0.05 & - \\
\hline
\end{tabular}
}
\tablefoot{Only an excerpt of the table is shown here. The full table is available at the CDS.}
\end{table*}

\subsection{Elemental abundances}
\label{sec:abundances}

To derive elemental abundances for species other than iron using NLTE grids in real-time (rather than through posterior corrections), only two spectral codes are currently available to our knowledge. These are the Python version of SME (Spectroscopy Made Easy, see \citealt{piskunov2017}), \texttt{pySME}, developed by \cite{wehr2023}, and the Python wrapper \texttt{TSFitPy} for Turbospectrum version 20 \citep{Plez2012}, as described in \citet{Gerber2023} and \citet{Storm2023}. Our chosen strategy was to adopt the latter option. \texttt{TSFitPy} allows the computation of synthetic spectra while incorporating NLTE corrections. The tool employs the Nelder-Mead algorithm to fit the normalized synthetic stellar spectra generated by Turbospectrum, minimizing the $\chi^2$ value in the process. The code offers access to two grids of stellar model atmospheres: the 1D line-blanketed hydrostatic MARCS models \citep{Gustafsson2008} and average 3D Stagger models \citep{Magic2013a, Magic2013b}. In this study, we chose to use the 1D MARCS stellar grid.

We determined NLTE abundances for 11 chemical elements: Mg, Si, Ti, Na, Al, Mn, Co, Ni, Sr, Y, and Eu. In principle, a few more elements could have been included; however, the Ca lines were saturated, the Ba abundances appeared unrealistically high, preventing reliable measurements. The line list was constructed through a careful selection of isolated, unblended, and sufficiently strong lines within our spectral range. Most lines originate from the Gaia-ESO Survey linelist, with Sr~{\sc i} 4607 \AA~ supplemented from VALD database\footnote{\tt \hyperlink{https://vald.astro.uu.se/}{https://vald.astro.uu.se/} }. For Eu, only the 6645 \AA~ line was used, as the strong resonance line at 4129 \AA~ is blended at the metallicity of our sample stars. We have included hyperfine splitting when needed (Mn, Co, and Eu lines). The complete line list is provided in Table \ref{tab:linelist_tsfit}.

The abundances were calculated as the average of the values derived for each line and the associated uncertainties were computed using $\sigma$/$\sqrt{n}$, where $\sigma$ is the standard deviation and n is the number of stars per cluster. For the cases when only one line was used, we adopted 0.05 dex as uncertainty, as it is the typical mean abundance uncertainty in our sample (see Table \ref{tab:sens} for sensitivities to stellar parameters). The chemical abundances for a subset of stars are presented in Table \ref{tab:abundances} (the full table is available at CDS \footnote{\tt \hyperlink{http://cdsweb.u-strasbg.fr/cgi-bin/qcat?J/A+A/}{http://cdsweb.u-strasbg.fr/cgi-bin/qcat?J/A+A/} }).

We established our abundance scale by analyzing a solar spectrum, using a spectrum of Ganymede obtained with the HARPS spectrograph on the 3.6 m ESO telescope. The stellar parameters adopted for the Sun are: $T_{\rm eff}$=5771 K, log $g$=4.44, logn(Fe)=7.52 and $\xi$=0.95 km~s$^{-1}$. The absolute solar abundances referenced by \texttt{TSFitPy} are sourced from \citet{Magg2022}. On the other hand, the determined solar abundances are as follows: A(Mg) = 7.47 $\pm$ 0.10, A(Si) = 7.49 $\pm$ 0.04, A(Ti) = 5.00 $\pm$ 0.03, A(Na) = 6.24 $\pm$ 0.01, A(Al) = 6.44 $\pm$ 0.02, A(Mn) = 5.48 $\pm$ 0.02, A(Co) = 4.84 $\pm$ 0.07, A(Ni) = 6.20 $\pm$ 0.03, A(Sr) = 2.72 $\pm$ 0.10, and A(Y) = 1.98 $\pm$ 0.04. Compared to the results of \citet{Asplund2009} and \citet{Magg2022}, our solar abundances are consistent within 2$\sigma$ uncertainties, except for Y, which exhibits a deviation of 4$\sigma$. The Eu~{\sc ii} line at 6645 \AA~was too weak in the solar spectrum to yield a reliable abundance measurement, and therefore no solar reference value could be determined. We adopted the value by \cite{Magg2022} of A(Eu)$_\odot$=0.52.  All other abundance ratios for our sample stars have been computed adopting the solar abundances derived in the present study.

\begin{table*}
\caption{Mean cluster abundance ratios.}
\label{tab:mean_abundances}
\setstretch{1.15}
\scalebox{0.6}{
\begin{tabular}{lrrrrrrrrrrrr}
\hline
\multicolumn{1}{c}{OC} & \multicolumn{1}{c}{[Fe/H]} & \multicolumn{1}{c}{[Mg/Fe]} & \multicolumn{1}{c}{[Si/Fe]} & \multicolumn{1}{c}{[Ti/Fe]} & \multicolumn{1}{c}{[Na/Fe]} & \multicolumn{1}{c}{[Al/Fe]} & \multicolumn{1}{c}{[Mn/Fe]} & \multicolumn{1}{c}{[Co/Fe]} & \multicolumn{1}{c}{[Ni/Fe]} & \multicolumn{1}{c}{[Sr/Fe]} & \multicolumn{1}{c}{[Y/Fe]} & \multicolumn{1}{c}{[Eu/Fe]} \\
\hline \hline                                                                      
Alessi 1         & -0.012 $\pm$ 0.01 & -0.025 $\pm$ 0.01 & 0.005  $\pm$ 0.02 & 0.056  $\pm$ 0.01 & 0.107  $\pm$ 0.01 & -0.027 $\pm$ 0.01 &  0.016 $\pm$ 0.01 & -0.061 $\pm$ 0.01 & 0.071  $\pm$ 0.01 & 0.020  $\pm$ 0.01 & 0.185 $\pm$ 0.01 & 0.098 $\pm$ 0.03 \\
AlessiTeutsch 11 & -0.040 $\pm$ 0.01 & -0.097 $\pm$ 0.05 & 0.014  $\pm$ 0.02 & 0.024  $\pm$ 0.02 & -0.066 $\pm$ 0.01 & -0.048 $\pm$ 0.02 & -0.086 $\pm$ 0.11 & -0.077 $\pm$ 0.06 & -0.122 $\pm$ 0.03 & -0.042 $\pm$ 0.05 & 0.166 $\pm$ 0.05 & - \\
Basel 11b        & 0.035  $\pm$ 0.02 & -0.078 $\pm$ 0.01 & -0.081 $\pm$ 0.01 & 0.012  $\pm$ 0.02 & 0.069  $\pm$ 0.05 & -0.015 $\pm$ 0.01 & -0.072 $\pm$ 0.02 & -0.090 $\pm$ 0.02 & -0.021 $\pm$ 0.02 & 0.109  $\pm$ 0.08 & 0.231 $\pm$ 0.05 & 0.166 $\pm$ 0.03 \\
COIN-Gaia 30     & -0.040 $\pm$ 0.01 & -0.069 $\pm$ 0.05 & -0.093 $\pm$ 0.02 & 0.012  $\pm$ 0.03 & 0.093  $\pm$ 0.01 & -0.047 $\pm$ 0.05 & -0.115 $\pm$ 0.09 & -0.098 $\pm$ 0.07 & -0.080 $\pm$ 0.05 & 0.119  $\pm$ 0.05 & 0.205 $\pm$ 0.05 & 0.132 $\pm$ 0.05 \\
Collinder 350    & -0.030 $\pm$ 0.08 & -0.085 $\pm$ 0.04 & -0.047 $\pm$ 0.05 & -0.047 $\pm$ 0.08 & 0.088  $\pm$ 0.04 & 0.020  $\pm$ 0.03 & -0.202 $\pm$ 0.13 & -0.096 $\pm$ 0.04 & -0.130 $\pm$ 0.03 & 0.138  $\pm$ 0.12 & 0.293 $\pm$ 0.10 & 0.204 $\pm$ 0.05 \\
Collinder 463    & -0.057 $\pm$ 0.01 & -0.083 $\pm$ 0.01 & -0.023 $\pm$ 0.01 & -0.015 $\pm$ 0.01 & 0.128  $\pm$ 0.01 & -0.034 $\pm$ 0.01 & -0.104 $\pm$ 0.01 & -0.102 $\pm$ 0.01 & -0.103 $\pm$ 0.01 & -0.055 $\pm$ 0.01 & 0.178 $\pm$ 0.02 & - \\
Gulliver 18      & -0.011 $\pm$ 0.02 & -0.061 $\pm$ 0.05 & -0.047 $\pm$ 0.03 & 0.061  $\pm$ 0.04 & 0.262  $\pm$ 0.06 & -0.023 $\pm$ 0.05 & - & -0.057 $\pm$ 0.07 & -0.192 $\pm$ 0.05 & 0.388  $\pm$ 0.05 & 0.393 $\pm$ 0.05 & 0.134 $\pm$ 0.04 \\
Gulliver 24      & -0.136 $\pm$ 0.02 & 0.020  $\pm$ 0.05 & -0.004 $\pm$ 0.03 & 0.115  $\pm$ 0.03 & 0.115  $\pm$ 0.01 & -0.001 $\pm$ 0.01 & -0.134 $\pm$ 0.13 & -0.021 $\pm$ 0.08 & -0.129 $\pm$ 0.04 & 0.219  $\pm$ 0.05 & 0.150 $\pm$ 0.12 & 0.168 $\pm$ 0.05 \\
Gulliver 51      & -0.170 $\pm$ 0.01 & -0.031 $\pm$ 0.05 & 0.111  $\pm$ 0.07 & 0.042  $\pm$ 0.06 & -0.015 $\pm$ 0.01 & -0.011 $\pm$ 0.01 & -0.061 $\pm$ 0.09 & -0.083 $\pm$ 0.06 & -0.055 $\pm$ 0.04 & -0.088 $\pm$ 0.05 & 0.104 $\pm$ 0.01 & 0.112 $\pm$ 0.05 \\
IC 4756          & -0.001 $\pm$ 0.01 & -0.031 $\pm$ 0.01 & -0.018 $\pm$ 0.01 & 0.112  $\pm$ 0.01 & 0.090  $\pm$ 0.01 & -0.049 $\pm$ 0.01 &  0.027 $\pm$ 0.01 & -0.048 $\pm$ 0.01 & 0.046  $\pm$ 0.01 & 0.093  $\pm$ 0.01 & 0.254 $\pm$ 0.01 & 0.134 $\pm$ 0.01 \\
NGC 2437         & -0.031 $\pm$ 0.04 & -0.103 $\pm$ 0.02 & -0.042 $\pm$ 0.02 & -0.037 $\pm$ 0.04 & 0.119  $\pm$ 0.02 & -0.038 $\pm$ 0.01 & -0.115 $\pm$ 0.08 & -0.191 $\pm$ 0.06 & -0.076 $\pm$ 0.06 & -0.111 $\pm$ 0.11 & 0.121 $\pm$ 0.05 & 0.096 $\pm$ 0.02 \\
NGC 2509         & 0.212  $\pm$ 0.01 & 0.020  $\pm$ 0.05 & 0.103  $\pm$ 0.05 & 0.163  $\pm$ 0.05 & 0.121  $\pm$ 0.07 & -0.032 $\pm$ 0.02 & - & 0.107  $\pm$ 0.06 & 0.157  $\pm$ 0.07 & -0.020 $\pm$ 0.10 & 0.282 $\pm$ 0.06 & - \\
NGC 2548         & 0.045  $\pm$ 0.03 & -0.058 $\pm$ 0.01 & -0.012 $\pm$ 0.04 & -0.030 $\pm$ 0.06 & 0.094  $\pm$ 0.04 & -0.061 $\pm$ 0.03 &  0.042 $\pm$ 0.00 & -0.154 $\pm$ 0.07 & 0.019  $\pm$ 0.02 & -0.010 $\pm$ 0.03 & 0.191 $\pm$ 0.06 & 0.050 $\pm$ 0.05 \\
NGC 2632         & 0.163  $\pm$ 0.03 & -0.049 $\pm$ 0.00 & 0.033  $\pm$ 0.01 & 0.049  $\pm$ 0.02 & 0.243  $\pm$ 0.04 & -0.055 $\pm$ 0.01 &  0.099 $\pm$ 0.01 & -0.045 $\pm$ 0.01 & 0.088  $\pm$ 0.02 & -0.116 $\pm$ 0.02 & 0.078 $\pm$ 0.03 & 0.039 $\pm$ 0.05 \\
NGC 2682         & 0.041  $\pm$ 0.01 & 0.025  $\pm$ 0.01 & 0.025  $\pm$ 0.01 & 0.142  $\pm$ 0.01 & 0.056  $\pm$ 0.01 & 0.007  $\pm$ 0.01 &  0.046 $\pm$ 0.04 & 0.016  $\pm$ 0.01 & 0.093  $\pm$ 0.01 & -0.088 $\pm$ 0.04 & 0.090 $\pm$ 0.01 & 0.036 $\pm$ 0.02 \\
NGC 6800         & 0.108  $\pm$ 0.01 & -0.091 $\pm$ 0.05 & -0.073 $\pm$ 0.04 & 0.081  $\pm$ 0.03 & 0.172  $\pm$ 0.01 & -0.029 $\pm$ 0.03 &  0.021 $\pm$ 0.09 & -0.087 $\pm$ 0.07 & 0.047  $\pm$ 0.06 & 0.043  $\pm$ 0.10 & 0.303 $\pm$ 0.06 & 0.080 $\pm$ 0.05 \\
NGC 6991         & 0.033  $\pm$ 0.03 & -0.055 $\pm$ 0.01 & 0.001  $\pm$ 0.01 & 0.117  $\pm$ 0.01 & 0.028  $\pm$ 0.01 & -0.055 $\pm$ 0.01 &  0.075 $\pm$ 0.01 & -0.048 $\pm$ 0.01 & 0.055  $\pm$ 0.01 & 0.058  $\pm$ 0.02 & 0.244 $\pm$ 0.02 & 0.075 $\pm$ 0.05 \\
NGC 7044         & -0.178 $\pm$ 0.04 & -0.086 $\pm$ 0.04 & -0.208 $\pm$ 0.01 & -0.062 $\pm$ 0.06 & 0.097  $\pm$ 0.10 & 0.025  $\pm$ 0.03 & - & -0.152 $\pm$ 0.05 & -0.245 $\pm$ 0.04 & 0.934  $\pm$ 0.06 & 0.252 $\pm$ 0.08 & - \\
NGC 7086         & -0.081 $\pm$ 0.02 & -0.087 $\pm$ 0.01 & -0.012 $\pm$ 0.02 & -0.023 $\pm$ 0.01 & 0.094  $\pm$ 0.01 & -0.014 $\pm$ 0.01 & -0.110 $\pm$ 0.00 & -0.095 $\pm$ 0.01 & -0.141 $\pm$ 0.02 & -0.060 $\pm$ 0.01 & 0.170 $\pm$ 0.01 & 0.154 $\pm$ 0.05 \\
NGC 7209         & -0.020 $\pm$ 0.05 & -0.072 $\pm$ 0.02 & -0.008 $\pm$ 0.04 & 0.023  $\pm$ 0.02 & 0.221  $\pm$ 0.03 & 0.005  $\pm$ 0.04 & -0.081 $\pm$ 0.05 & -0.053 $\pm$ 0.03 & -0.076 $\pm$ 0.05 & 0.027  $\pm$ 0.06 & 0.207 $\pm$ 0.07 & - \\
NGC 752          & 0.023  $\pm$ 0.06 & -0.052 $\pm$ 0.02 & 0.011  $\pm$ 0.03 & 0.111  $\pm$ 0.02 & 0.004 $\pm$ 0.02 & -0.068 $\pm$ 0.02 &  0.049 $\pm$ 0.02 & -0.040 $\pm$ 0.02 & 0.047  $\pm$ 0.02 & 0.059  $\pm$ 0.03 & 0.216 $\pm$ 0.02 & 0.085 $\pm$ 0.01 \\
Ruprecht 171     & -0.075 $\pm$ 0.02 & -0.006 $\pm$ 0.02 & 0.032  $\pm$ 0.04 & 0.047  $\pm$ 0.02 & 0.049  $\pm$ 0.02 & 0.021  $\pm$ 0.02 & -0.067 $\pm$ 0.05 & -0.030 $\pm$ 0.01 & -0.026 $\pm$ 0.03 & 0.177  $\pm$ 0.15 & 0.142 $\pm$ 0.06 & 0.078 $\pm$ 0.02 \\
Stock 2          & -0.019 $\pm$ 0.02 & -0.101 $\pm$ 0.02 & -0.081 $\pm$ 0.03 & 0.001  $\pm$ 0.01 & 0.089  $\pm$ 0.01 & -0.039 $\pm$ 0.01 & -0.030 $\pm$ 0.01 & -0.114 $\pm$ 0.02 & -0.050 $\pm$ 0.03 & 0.027  $\pm$ 0.01 & 0.209 $\pm$ 0.02 & 0.104 $\pm$ 0.05 \\
Tombaugh 5       & 0.016  $\pm$ 0.04 & -0.004 $\pm$ 0.04 & -0.047 $\pm$ 0.05 & -0.039 $\pm$ 0.06 & 0.133  $\pm$ 0.06 & 0.002  $\pm$ 0.04 & - & -0.042 $\pm$ 0.05 & -0.088 $\pm$ 0.08 & - & 0.234 $\pm$ 0.03 & 0.170 $\pm$ 0.01 \\
UBC 3            & 0.005  $\pm$ 0.01 & -0.050 $\pm$ 0.05 & -0.058 $\pm$ 0.02 & 0.038  $\pm$ 0.03 & 0.097  $\pm$ 0.01 & 0.052  $\pm$ 0.05 &  0.028 $\pm$ 0.01 & -0.087 $\pm$ 0.06 & -0.065 $\pm$ 0.05 & -0.024 $\pm$ 0.05 & 0.205 $\pm$ 0.05 & 0.184 $\pm$ 0.05 \\
UBC 60           & 0.227  $\pm$ 0.01 & -0.066 $\pm$ 0.05 & -0.037 $\pm$ 0.05 & 0.110  $\pm$ 0.04 & 0.154  $\pm$ 0.02 & -0.037 $\pm$ 0.03 &  0.159 $\pm$ 0.09 & -0.055 $\pm$ 0.08 & 0.056  $\pm$ 0.05 & 0.005  $\pm$ 0.05 & 0.266 $\pm$ 0.05 & - \\
UBC 131          & 0.052  $\pm$ 0.02 & -0.014 $\pm$ 0.01 & -0.020 $\pm$ 0.01 & 0.121  $\pm$ 0.01 & 0.055  $\pm$ 0.01 & -0.054 $\pm$ 0.01 & - & -0.054 $\pm$ 0.01 & 0.130  $\pm$ 0.01 & 0.095  $\pm$ 0.02 & 0.269 $\pm$ 0.02 & - \\
UBC 141          & -0.023 $\pm$ 0.01 & 0.079  $\pm$ 0.05 & -0.015 $\pm$ 0.03 & 0.165  $\pm$ 0.03 & -0.021 $\pm$ 0.01 & -0.047 $\pm$ 0.01 & -0.010 $\pm$ 0.07 & -0.014 $\pm$ 0.07 & 0.061  $\pm$ 0.06 & 0.037  $\pm$ 0.05 & 0.189 $\pm$ 0.04 & - \\
UBC 169          & 0.053  $\pm$ 0.02 & -0.216 $\pm$ 0.05 & -0.244 $\pm$ 0.03 & -0.084 $\pm$ 0.04 & 0.062  $\pm$ 0.03 & -0.053 $\pm$ 0.04 & - & -0.173 $\pm$ 0.09 & -0.218 $\pm$ 0.02 & -0.139 $\pm$ 0.05 & 0.103 $\pm$ 0.05 & - \\
UBC 170          & 0.091  $\pm$ 0.01 & -0.125 $\pm$ 0.03 & -0.119 $\pm$ 0.03 & -0.063 $\pm$ 0.04 & 0.144  $\pm$ 0.02 & -0.071 $\pm$ 0.01 & -0.109 $\pm$ 0.09 & -0.222 $\pm$ 0.08 & -0.127 $\pm$ 0.06 & -0.097 $\pm$ 0.03 & 0.203 $\pm$ 0.03 & - \\
UBC 194          & 0.055  $\pm$ 0.01 & - & - & - & - & -0.148 $\pm$ 0.05 &  0.016 $\pm$ 0.01 & - & -0.355 $\pm$ 0.07 & - & - & - \\
UBC 577          & -0.033 $\pm$ 0.02 & -0.030 $\pm$ 0.01 & 0.020  $\pm$ 0.02 & 0.106  $\pm$ 0.01 & 0.074  $\pm$ 0.00 & 0.017  $\pm$ 0.03 &  0.175 $\pm$ 0.08 & -0.054 $\pm$ 0.02 & 0.065  $\pm$ 0.01 & 0.069  $\pm$ 0.01 & 0.239 $\pm$ 0.01 & - \\
UPK 219          & 0.070  $\pm$ 0.01 & -0.124 $\pm$ 0.05 & -0.125 $\pm$ 0.02 & -0.032 $\pm$ 0.02 & 0.091  $\pm$ 0.02 & -0.018 $\pm$ 0.03 & -0.091 $\pm$ 0.09 & -0.118 $\pm$ 0.06 & -0.079 $\pm$ 0.05 & -0.019 $\pm$ 0.05 & 0.170 $\pm$ 0.05 & 0.120 $\pm$ 0.05 \\
\hline
\end{tabular}
}
\end{table*}

We checked for correlations of [X/H] ratios with temperature and surface gravity considering the same open clusters in Fig.~\ref{fig:lotus}, and the results are shown in Fig.~\ref{fig:abund_trends}. Within uncertainties, we find no significant trends for Mg, Si, Ti, Na, Al, as their behavior remains generally flat. However, for Mn, Co, Ni, Sr, Y and Eu, the abundances increase slightly with temperature (and $\log g$) and there is also notable scatter. Moreover, for Sr and Y we have detected an upturn for $T_{\rm eff} \lesssim 4500$ K. 

Table \ref{tab:mean_abundances} shows the mean abundances for the 33 OCs. The abundance of each cluster is the average considering its members and the error is given by $\sigma$/$\sqrt{n}$. For clusters with only one star, the uncertainty is the standard deviation. Among our clusters, some interesting cases exist, such as the super-solar metallicity OCs NGC 2509 with [Fe/H]=0.212 $\pm$ 0.01 and UBC 60 with [Fe/H]=0.227 $\pm$ 0.01. On the other hand, Gulliver 51 and NGC 7044 exhibit the lowest values of our sample, with [Fe/H]$\sim$-0.2 dex.

Overall, we confirm that all the open clusters display a solar-scaled chemical composition concerning the $\alpha$ elements, specifically Mg, Si, and Ti and the iron peak elements Mn, Co, and Ni. While Na abundances are generally consistent with solar values within the uncertainties, they display the well-known trend correlated with age. Neutron-capture element Sr and Eu show a solar-scaled abundance pattern, whereas Y is slightly super-solar. Regarding the potential enhancement of s-process elements, which may be associated with magnetic activity at younger ages, we direct readers to \cite{Spina2018}, \cite{baratella2021}, \cite{nordlander2024}, and the references therein for further discussion.


\section{Comparison with the literature}
\label{sec:comp_lit}

We used the open clusters IC~4756, NGC~752, NGC~6991 and UBC~3, which have been previously studied in the literature, for comparison with our results. Furthermore, as explained before, we incorporate open clusters previously analyzed by \citet{Casali2020a}, \citet{Dorazi2020}, \citet{Zhang2021, Zhang2022}, and \citet{Alonso2021}. Some clusters studied by these authors have also been investigated in the literature through other independent studies. Here, we present a comprehensive comparison of all available data, integrating results from both our study and previous works to ensure consistency and identify any discrepancies. In this process, we limited to high-resolution, optical spectroscopy. The comparison results are summarized in Table \ref{tab:comparison_literature} and Fig. \ref{fig:parameters_atm_stars}. 

For $T_{\rm eff}$ and log $g$, we compared our results with the literature both on a cluster-by-cluster basis, by calculating the mean difference (this work minus literature) for each OC using stars in common, and on a star-by-star basis for individual objects.
Considering the four OCs that are new in the SPA framework, we identified in the literature eight common stars in IC 4756 (\#0, \#2, \#4, \#5, \#7, \#12, \#13, \#14), three in NGC 752 (\#3, \#4, \#5), one in NGC 6991 (\#1) and one in UBC 3 (\#1) (see Table \ref{tab:comparison_literature} and Fig. \ref{fig:parameters_atm_stars} for the references). 

In general, there is a good agreement in the results, considering the mean differences in $T_{\rm eff}$ and log $g$, as shown in Table \ref{tab:comparison_literature} and Fig. \ref{fig:parameters_atm_stars}. For [Fe/H] and [Mg/Fe] our results show an overall agreement, while [Si/Fe] is generally lower in this work. [Ti/Fe], [Na/Fe] and [Al/Fe] display some variations, typically up to 0.2 dex, with aluminium reaching -0.29 dex in NGC 752. [Mn/Fe] is generally consistent, with a maximum difference of 0.3 dex in NGC 752. [Co/Fe] and [Ni/Fe] are in good agreement, with differences up to 0.15 dex. For [Sr/Fe], with fewer clusters available for comparison, the maximum difference is -0.26 dex in UBC 3, while [Y/Fe] shows differences of around 0.2 dex. Similarly, [Eu/Fe] is available for a limited number of clusters, with a maximum difference of 0.18 dex. Notably, \citet{Jacobson2007} shows considerable variations for [Si/Fe], [Na/Fe], and [Al/Fe] in IC 4756 compared to our results. Note that the works presented here may use different solar scales, which could explain some of the observed differences. 

\begin{figure}
\centering
\includegraphics[width=\linewidth]{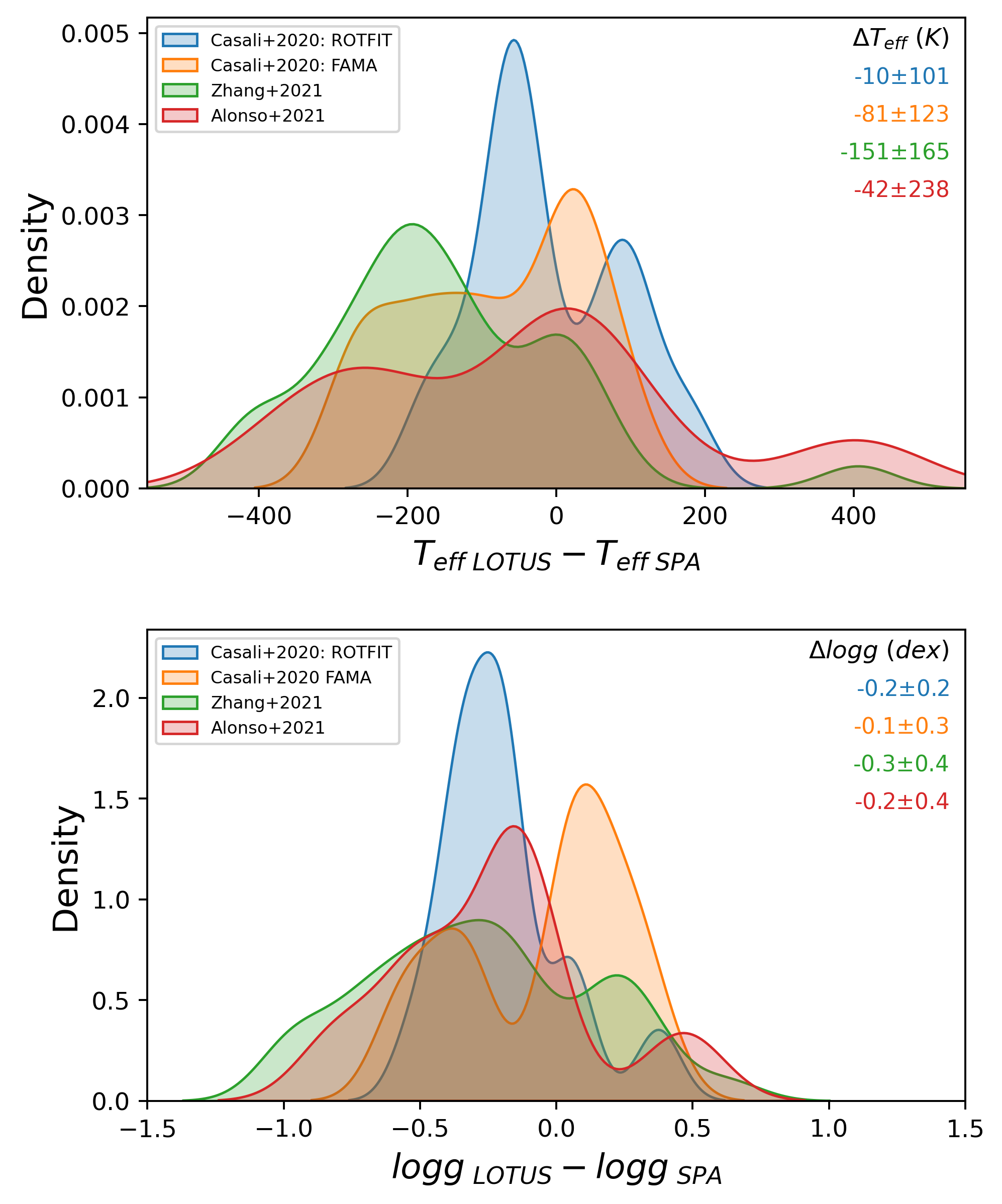}
\caption{Distribution of $\Delta T_{\rm eff}$ and $\Delta \log g$ comparing LOTUS results to those from previous SPA studies. The histograms were smoothed. The colors indicate the different authors. Mean differences and standard deviations are displayed in each panel.}
\label{fig:param_diff}
\end{figure}

Considering the SPA OCs previously published, \citet{Casali2020a} presented the spectral analysis of 15 stars in four OCs: Collinder 350, Gulliver 51, NGC 7044, and Ruprecht 171. The authors derived stellar parameters using both EW analysis with the Fast Automatic {\tt MOOG} Analysis ({\tt FAMA}, \citet{Magrini2013}) and a spectral fitting technique with {\tt ROTFIT} \citep{Frasca2006, Frasca2019}. One of the stars in Gulliver 51 ($\#2$), identified as a fast rotator, was excluded from our analysis. The $T_{\rm eff}$ and log $g$ show a good agreement, as indicated by their mean differences (see Table \ref{tab:comparison_literature}). More recently, \citet{CarbajoHijarrubia2024} also investigated Ruprecht 171, revealing three stars in common with our study ($\#5$, $\#6$, and $\#8$), which show general agreement, with a mean difference up to 220 K in temperature and 0.22 dex in [Fe/H]. 

\citet{Zhang2021} derived stellar parameters for 40 giant stars in 18 open clusters using EW measurements. From the initial set, we excluded six binary stars (Gulliver 37 \#1, NGC 2437 \#6, NGC 2548 \#4, NGC 2682 \#3 and \#4, NGC 7082 \#1 and \#2) along with two stars (ASCC 11 \#1 and NGC 2548 \#4) that did not reach convergence with {\tt LOTUS}. The results demonstrate good agreement in the mean differences of $T_{\rm eff}$ and log $g$, with the exceptions of COIN-Gaia 30 and Gulliver 18. Furthermore, other authors have studied some of these OCs. We identified three stars in common in NGC 2548 (\#1, \#2, and \#3), two stars in NGC 2682 (\#1 and \#2), and one star in Tombaugh 5 (\#1), see Table \ref{tab:comparison_literature} for the references. All of them demonstrate overall consistency in $T_{\rm eff}$ and log $g$ with our results. Regarding the abundance measurements, Table \ref{tab:comparison_literature} indicates agreement for [Fe/H] and [Mg/Fe], while our results generally show lower [Si/Fe] values. Differences in [Ti/Fe] are observed, typically up to 0.2 dex, with a maximum of 0.28 dex for NGC 2682. Additionally, [Na/Fe], [Al/Fe], [Mn/Fe], [Co/Fe], and [Ni/Fe] exhibit agreement within 0.2 dex. Finally, [Y/Fe] values in our study are generally higher compared to the literature, reaching 0.26 for NGC 2548. We need however to consider that generally LTE results are reported.

Finally, \citet{Alonso2021} analyzed 46 stars (both dwarfs and giants) in the open cluster Stock 2, determining the stellar atmospheric parameters using {\tt ROTFIT} and abundances using {\tt SYNTHE} and an optimization tool. In our analysis, we have 10 stars in common with \citet{Alonso2021}, although we excluded two binary stars (Stock 2 g3 and g5) from our sample. Additionally, as noted by \citet{Alonso2021}, Stock 2 was also studied by \citet{Reddy2019}, with which we have two stars in common (\#4 and \#9). In general, the results are in fair agreement.  

Figure \ref{fig:param_diff} shows the smoothed distribution of the differences in $T_{\rm eff}$ and $\log g$ for the stars in common considering this work and previous SPA studies. The distributions of $\Delta T_{\rm eff}$ and $\Delta$ log $g$ reveal significant spreads, particularly for \citet{Zhang2021} and \citet{Alonso2021}, which appear broader compared to \citet{Casali2020a} ({\tt ROTFIT} and {\tt FAMA}). Notably, however, the majority of stars display differences concentrated around zero, indicating reasonable overall agreement between the results. The significant spreads in these distributions are primarily driven by a small subset of outlier stars. The mean values further suggest that {\tt LOTUS} tends to yield lower values of $T_{\rm eff}$ and $\log g$ compared to previous SPA results, as indicated by the leftward peaks in the distributions.

\begin{figure}
\centering
\includegraphics[width=\linewidth]{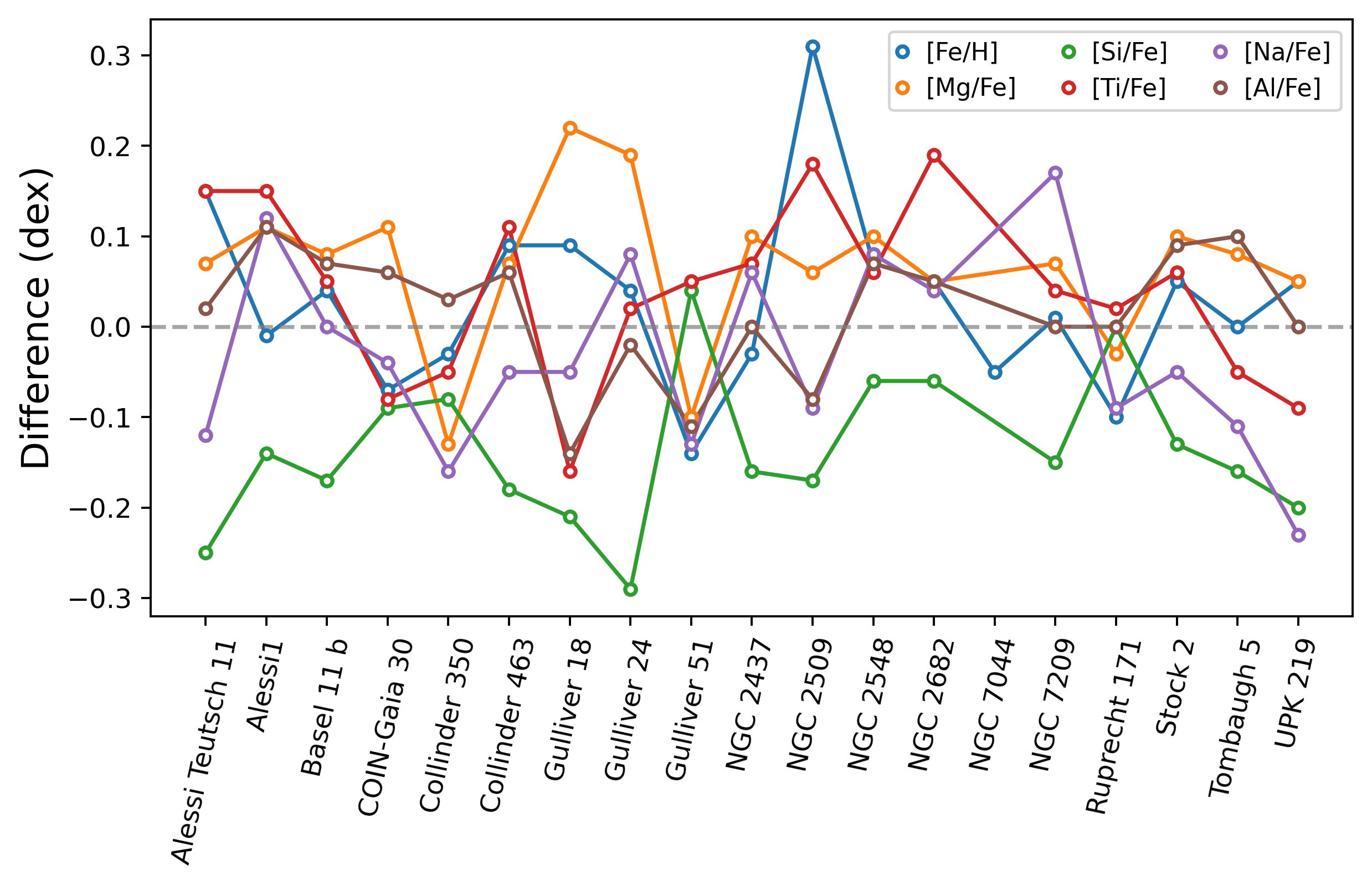}
\caption{Comparison between our abundance of Fe, $\alpha$ and odd-z elements results and those from previous SPA studies for each open cluster. The lines represent different elements. }
\label{fig:abund_diff}
\end{figure}

Figure \ref{fig:abund_diff} presents a comparative analysis of our results alongside previous SPA studies, focusing solely on SPA because it uses the same spectral data while employing different methodologies for calculating abundances. Notably, \citet{Zhang2022} applied NLTE corrections specifically to Na and Al, whereas \citet{Casali2020a} addressed NLTE effects for Fe but considered them negligible. Overall, we identify 19 SPA OCs in common for this comparison. Each line in the figure represents a distinct abundance, with our focus limited to [Fe/H], $\alpha$, and odd-Z elements. The comparison emphasizes several key points: i) The differences in [Fe/H] values are generally close to 0.0 dex, indicating strong consistency, except for the case of NGC 2509; ii) The [Mg/Fe] ratio displays more variation, with a few clusters deviating by approximately $\pm$0.2 dex; iii) [Si/Fe] results are slightly lower in the present analysis, particularly for Alessi Teutsch 11, Gulliver 18, Gulliver 24, and UPK 219; iv) Differences in [Ti/Fe] are present, with maximum value reaching around $\pm$0.15 dex; v) [Na/Fe] exhibits substantial discrepancy, approaching $\pm$0.2 dex; vi) [Al/Fe] shows similar trends to [Na/Fe], with variations present but generally less pronounced than those for Na. Overall, the comparison highlights that, despite some variations, the results presented here align well with previous studies for most elements. The observed differences in the comparison can primarily be attributed to three factors: the use of different line lists, variations in spectral analysis methodologies, and the adoption of NLTE abundances instead of LTE or a posteriori NLTE corrections. The methods used in this study resulted in reduced star-to-star scatter within each cluster, thereby providing higher internal precision.


\begin{figure*}
\centering
\includegraphics[width=\linewidth]{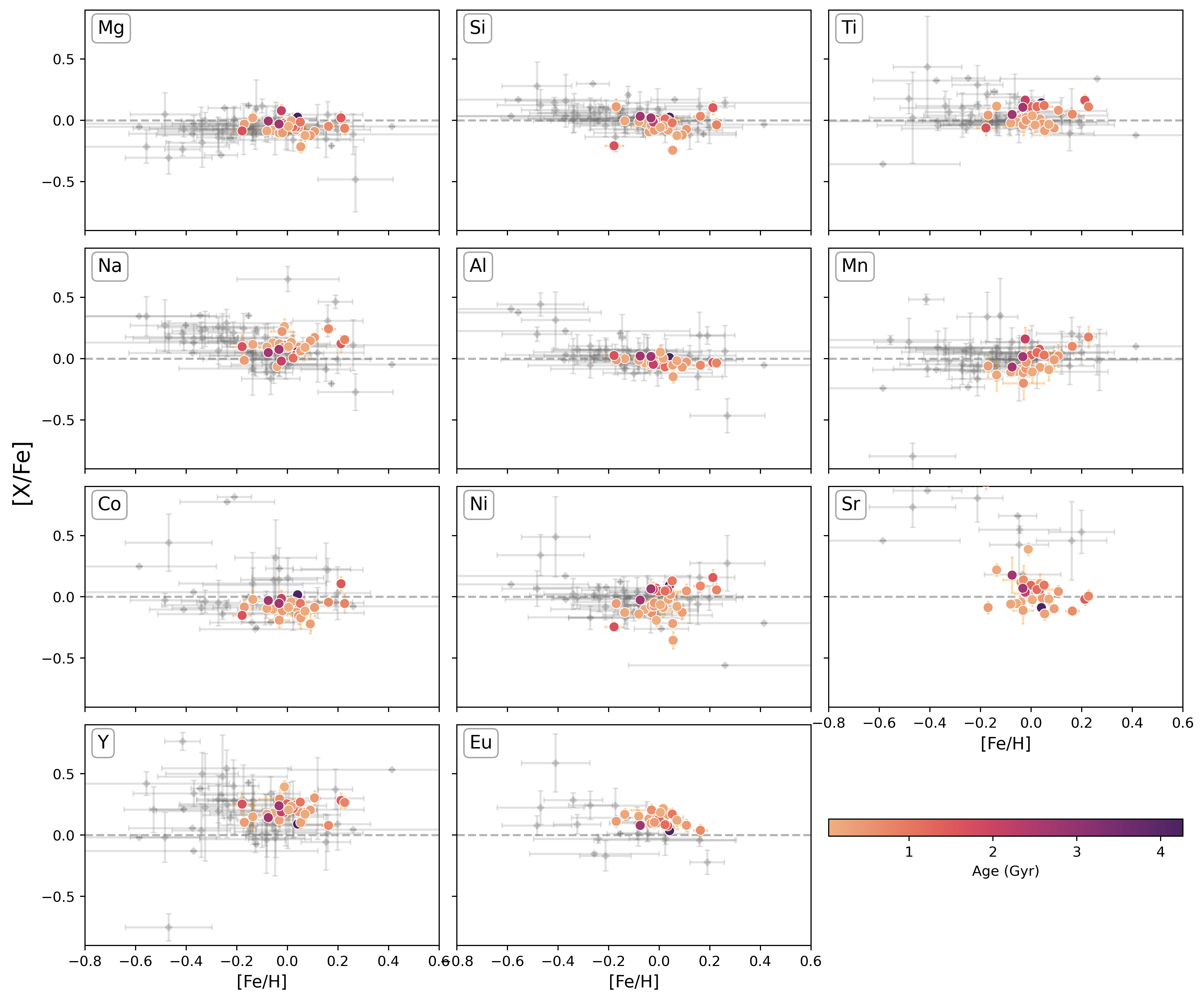}
\caption{Abundance [X/Fe] ratios as function of [Fe/H]. The gray diamonds represent the OCs compilation from GALAH (see Table~\ref{tab:galah_oc}). Our results are color-coded according to the age.}
\label{fig:elem_fe}
\end{figure*}

\begin{figure}
\centering
\includegraphics[width=\linewidth]{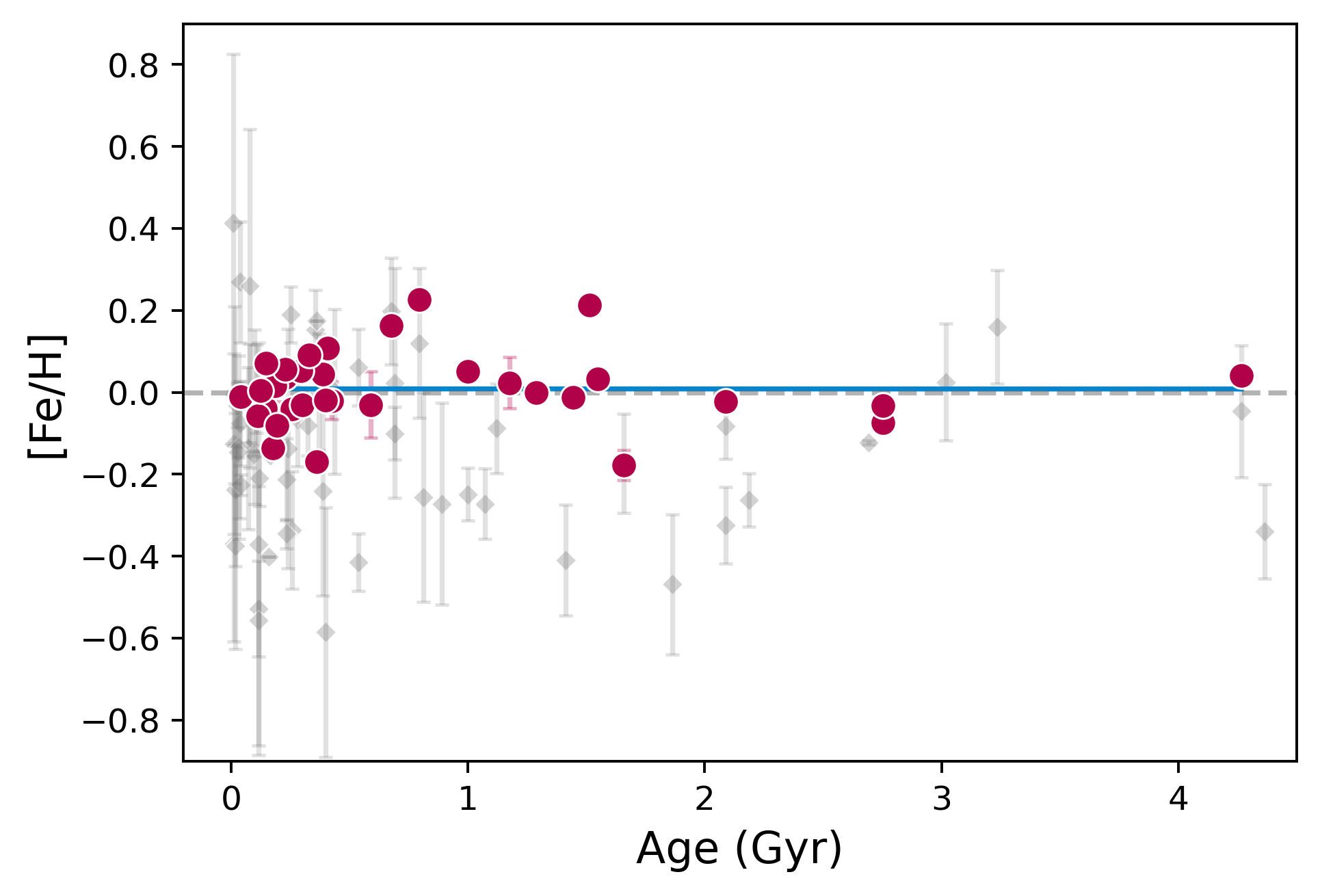}
\caption{Mean [Fe/H] per cluster as a function of age. The red points represent our sample of 33 open clusters, while the gray diamonds correspond to clusters from GALAH.}
\label{fig:iron_age}
\end{figure}

\section{Galactic trends}\label{sec:gal_trends}

We derived abundances for Fe and 11 chemical elements that are synthesized in different nucleosynthesis sites, including Type II and Type Ia supernovae (SNe) and asymptotic giant branch stars (AGB), each contributing to the interstellar medium at different rates (see, e.g., \citealt{Kobayashi2020}; see also \citealt{Romano2010}, where uncertainties due to imperfect knowledge of stellar yields are discussed). More in general, studying how abundances change over time and across the Galactic disk is crucial for understanding the channels of element production and the various processes that have shaped the Milky Way's history \citep{molero2023}. In this context, OCs present a favorable option, as they are distributed throughout the Galactic disk and encompass a wide range in ages that reflects a significant fraction of the lifespan of the thin disk. Furthermore, it is possible to characterize the chemistry of an OC using a small subset of its members. 

We explored the abundance trends in the Galactic disk using the mean abundances for the 33 open clusters, combined with ages from \citet{Cantat2020}. Our sample does not cover a large range in $R_{\textrm{GC}}$. Most of our clusters are within about 1 kpc from the Sun (see Table~\ref{tab:lit_sum}) and farther away from the Galactic center than the Sun, reflecting the fact that we observed from the Northern hemisphere. Conversely, our sample has broad coverage in age, from about 40 Myr to about 4 Gyr. We then opted to concentrate on the (possible) variations and evolution of abundances with age. 

\begin{figure*}
\centering
\includegraphics[width=\linewidth]{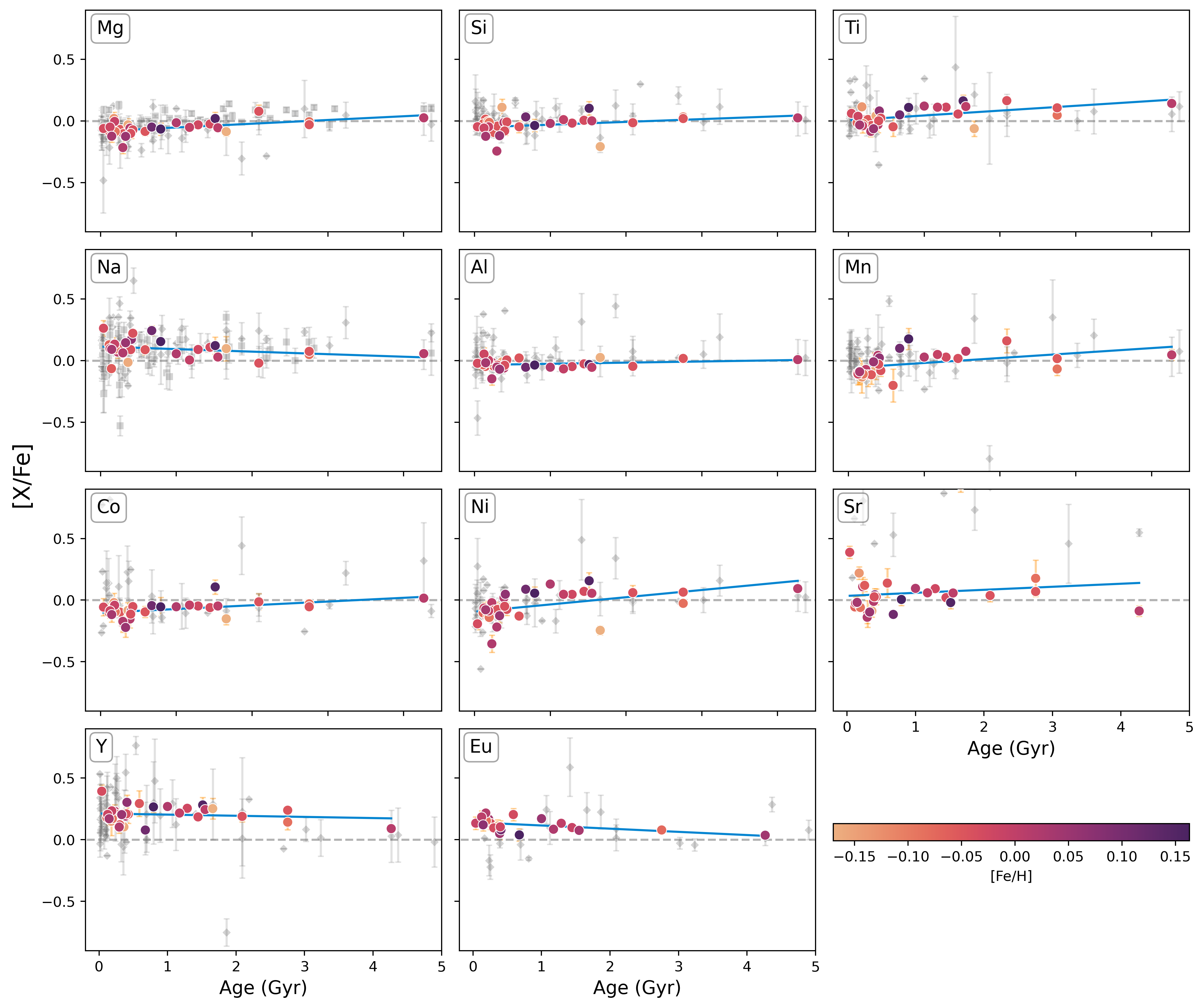}
\caption{Abundance [X/Fe] ratios as a function of age. The gray diamonds represent the OCs compilation from GALAH (see Table~\ref{tab:galah_oc}) and the gray squares represent the OCs from OCCAM. Our results are color-coded according to [Fe/H].} 
\label{fig:elem_age_fe}
\end{figure*}

We included the OC sample from \citet{Spina2021} for visual comparison only, focusing specifically on the NLTE abundances reported in GALAH DR3 \citep{Buder2021}. Since our analysis is centered on NLTE abundances, we excluded comparisons with clusters from the Gaia-ESO  survey \citep[see, e.g.,][]{Magrini2023}. Instead, Mg and Na from APOGEE DR17 were compared,  as these elements were derived using NLTE corrections. They were included using the sample from \citet{Myers2022}. Although Spina et al. standardized data from APOGEE and GALAH, our calculations of mean abundances were based exclusively on GALAH. The ages presented in the GALAH and APOGEE OC compilation are taken from \citet{Cantat2020}. Our discussion encompasses a total of 63 OCs from the original sample reported by \citet{Spina2021}, which are listed in Table \ref{tab:galah_oc}, as well as an additional 84 OCs for Mg and 64 OCs for Na from \citet{Myers2022}. The GALAH and the APOGEE OCs are represented by gray symbols in Figs. \ref{fig:elem_fe}, \ref{fig:iron_age}, and \ref{fig:elem_age_fe}. 

The elements in the diagrams can be divided into categories: $\alpha$ elements (Mg, Si, and Ti), iron-peak elements (Mn, Co, and Ni), odd-Z elements (Na and Al), and neutron-capture elements (Sr, Y, and Eu). Overall, our findings are consistent with the distribution observed in GALAH, except for the [Sr/Fe] ratio; however, our abundance ratios show notably less scatter (see Fig.~\ref{fig:elem_fe} and details in Sect. \ref{sec:neutron_capture}). 

Regarding the age dependence, Fig.~\ref{fig:iron_age} illustrates the trend of [Fe/H] with age. As anticipated, there is no significant trend in OC metallicity relative to age, as the slope of the [Fe/H] versus age data is negligible. This supports the formation locus of a cluster as a more important factor for its chemical composition than the age—suggesting a lack of an age-metallicity relationship. Yet, it must be noted that only a mild evolution is expected in the last 4.5 Gyr in the solar neighborhood. It is important to mention that when focusing on very young ages, most (though not all) of the clusters show slightly subsolar iron abundances \citep{kos2021, Alonso2024}. This observation appears to challenge the conventional model of Galactic chemical evolution, unless there has been a recent inflow of metal-poor material (see, e.g., \citealt{Spitoni2023, palla2024}). This situation is not typical, and we would anticipate different behavior based on modeling predictions. However, magnetic activity may indeed be influencing the results \citep{Spina2020, baratella2021}.

For each of the [X/Fe] ratios we performed a linear fit of the abundance distributions versus age (shown in Fig.~\ref{fig:elem_age_fe} as a thick blue line). The corresponding slopes and their uncertainties are given in Table~\ref{tab:gradients}, alongside the values for the SPA sample from \citet{Zhang2022} (18 OCs) and the OCCASO and OCCAM projects \citep{CarbajoHijarrubia2024, Myers2022}, which include 26 and 84 OCs, respectively. Since \citet{Myers2022} did not provide slopes for [X/Fe] versus age, we derived them. The slopes obtained in this work are compatible within the errors with those used for comparison. For Ti, Mn, and Ni, our slopes (0.039 $\pm$ 0.011, 0.040 $\pm$ 0.016, and 0.059 $\pm$ 0.018, respectively) are steeper than those from other studies, which range from 0.002 to 0.012 for Ti, -0.008 to 0.003 for Mn, and 0.010 to 0.026 for Ni. In OCCAM, a positive slope is found for Na (0.004 $\pm$ 0.009), whereas this work (-0.021 $\pm$ 0.012), the previous SPA study, and OCCASO are in agreement with negative slopes. Yttrium has a shallower slope (-0.009 $\pm$ 0.013) in this work compared to OCCASO (-0.022 $\pm$ 0.012). No comparison is provided for Eu, as none of the references report a corresponding slope.

\subsection{$\alpha$ elements}

The $\alpha$ elements are mainly produced by core-collapse supernovae (SNe), marking the final stage of the most massive stars (M $\geq$ 10 $\rm M_{\odot}$). Since these massive stars have short lifetimes ($\leq 10^{-2}$ Gyr), they quickly enrich the interstellar medium when compared with Fe, which is mainly produced on longer timescales via Type Ia SNe. In Fig.~\ref{fig:elem_fe}, Mg, Si and Ti shows a relatively flat behavior with increasing metallicity, consistent with observations from GALAH open clusters; Mg is also consistent with OCCAM open clusters. Figure \ref{fig:elem_age_fe} shows a decrease in the abundance ratios of Mg, Si, and Ti with younger ages. This trend mirrors observations made for nearby solar twins \citep{Nissen2015, Spina2016, Casali2020b} and field stars \citep[see, e.g.][]{Buder2019, DelgadoMena2019, Hayden2022}. Such behavior is anticipated in the Galactic disk, which was initially formed from gas that was pre-enriched with $\alpha$ elements, followed by additional enrichment from subsequent generations of stars. The observed decline in [$\alpha$/Fe] ratios with younger ages is consistent with predictions from Galactic chemical evolution models. In these models, successive generations of stars significantly contribute to $\alpha$ element enrichment through core-collapse supernovae, while the delayed contributions from Type Ia supernovae dilute more and more efficiently the ejecta from core-collapse supernovae as the time goes by.

\subsection{Odd-z elements}

Sodium and aluminum are primarily generated by core-collapse supernovae, but they are also produced during the AGB phase (see \citet{Smiljanic2016}). In AGB models, Na production is notably affected by metallicity, exhibiting secondary behavior where stars with higher metallicities tend to have larger Na abundances. Figure \ref{fig:elem_fe} displays a generally flat trend for sodium in our sample, despite some scatter. While a distinct declining trend is observed in the GALAH open clusters, which demonstrate a wider range of metallicities compared to our sample, we must recognize the limitations posed by the smaller size of the SPA sample. This limitation introduces uncertainty regarding the interpretation of our findings, especially concerning the flatness and observed upward trend in [Na/Fe] for [Fe/H] > 0, which may be compatible with the GALAH distribution. This observation is consistent with recent findings by \cite{owusu2024} and aligns with previous studies by e.g., \citealt{adibekyan2012, Nissen2020}. In contrast to Na, the [Al/Fe] trends are more closely associated with those of the $\alpha$ elements, likely due to their shared nucleosynthetic origins in massive stars (the AGB contribution to [Al/Fe] should be marginal in this metallicity range). 

Figure \ref{fig:elem_age_fe} reveals several age-dependent trends, with Na decreasing and Al showing a mild increase as age progresses. The correlation between the behavior of Al and the $\alpha$ elements over time has also been observed in solar twins \citep{Casali2020b} and field stars \citep{DelgadoMena2019}. The observed trend of decreasing [Na/Fe] with age is consistent with what was found, for example, by \citet{Smiljanic2016, Smiljanic2018}. They observed an increase in the surface Na abundance for red giants belonging to OCs with turnoff mass larger than about 2 M$_\odot$ (i.e., younger than about 1 Gyr) and for Cepheids. This was attributed to evolutionary mixing processes; for these masses, additional Na is brought to the star surface at the first dredge-up, as confirmed by stellar evolutionary models predictions (see e.g., \citealt{Lagarde2012, Ventura2013}).

\begin{table}
\caption{[X/Fe] versus age slopes and associated uncertainties.}
\label{tab:gradients}
\centering
\setstretch{1.15}
\scalebox{0.7}{
\begin{tabular}{lrrrr}
\hline
\multicolumn{1}{c}{Element} & \multicolumn{1}{c}{This work} & \multicolumn{1}{c}{SPA} & \multicolumn{1}{c}{OCCASO} &  \multicolumn{1}{c}{OCCAM} \\
\hline \hline
Mg &  0.030 $\pm$ 0.008 &  0.036 $\pm$ 0.007 &  0.009 $\pm$ 0.007 & 0.022 $\pm$ 0.002 \\
Si &  0.023 $\pm$ 0.013 & -0.014 $\pm$ 0.006 &  0.012 $\pm$ 0.008 & 0.011 $\pm$ 0.003 \\
Ti &  0.039 $\pm$ 0.011 &  0.002 $\pm$ 0.010 &  0.009 $\pm$ 0.005 & 0.012 $\pm$ 0.003 \\ 
Na & -0.021 $\pm$ 0.012 & -0.031 $\pm$ 0.040 & -0.030 $\pm$ 0.013 & 0.004 $\pm$ 0.009 \\
Al &  0.009 $\pm$ 0.006 &  0.009 $\pm$ 0.008 &  0.020 $\pm$ 0.011 & 0.026 $\pm$ 0.004 \\
Mn &  0.040 $\pm$ 0.016 &  - &  0.003 $\pm$ 0.007 & -0.008 $\pm$ 0.003\\
Co &  0.030 $\pm$ 0.010 &  - &  0.007 $\pm$ 0.005 & 0.016 $\pm$ 0.008 \\
Ni &  0.059 $\pm$ 0.018 &  - &  0.026 $\pm$ 0.006 & 0.010 $\pm$ 0.002 \\
Sr &  0.025 $\pm$ 0.036 &  - & -0.070 $\pm$ 0.008 &  -  \\
Y  & -0.009 $\pm$ 0.013 &  - & -0.022 $\pm$ 0.012 &  - \\
Eu & -0.026 $\pm$ 0.010 &  - & - &  - \\
\hline
\end{tabular}
}
\tablefoot{The SPA slopes correspond to the sample of 18 open clusters presented in  \citet{Zhang2022}. For OCCASO, they correspond to 36 open clusters in the recent work of \citet{CarbajoHijarrubia2024}. Values for [X/Fe] ratios and ages come from the original papers. For OCCAM, it is based on 84 high-quality open clusters from APOGEE DR17 \citet{Myers2022}, except for Na (64), Al and Mn (82), and Co (61).}
\end{table}

\subsection{Iron-peak elements}

Type Ia SNe, resulting from white dwarf stars in interacting binary systems \citep{whelan1973, iben1984}, are the primary source of iron-peak elements \citep{kobayashi2011, Kobayashi2020, nomoto2018}. 
Both core-collapse supernovae (CCSNe) and Type Ia supernovae contribute to the nucleosynthesis of iron-group elements, including manganese. For metallicities of [Fe/H] $ \lesssim -1$, CCSNe dominate Mn production in the solar neighborhood. However, the transition point where Type Ia SNe begin to dominate varies across the Galactic disc: it occurs at lower [Fe/H] in the inner disc, where chemical evolution proceeds more rapidly, and at higher [Fe/H] in the outer disc, which evolves more slowly \citep{Seitenzahl2013}. Despite a general consensus on the qualitative roles of CCSNe and SNe Ia in producing Mn and Fe, the quantitative contributions remain uncertain. Consequently, the observed evolution of the [Mn/Fe] ratio with metallicity across the Galaxy is still not fully understood. \cite{eitner2020} demonstrated that after accounting for NLTE corrections, the [Mn/Fe] ratio remains solar across all metallicities, from approximately [Fe/H] $\approx -4$ to 0. Our sample enables us to explore the chemical evolution of Mn at super-solar metallicities. As shown in Fig.~\ref{fig:elem_fe}, we observe an increasing trend in the [Mn/Fe] ratios with rising metallicity. 

The [Co/Fe] ratios seem to follow a quite flat trend, in agreement with previous studies for thin disk stars (e.g., \citealt{reddy2003, reddy2006}). On the other hand, we detected a mild decreasing trend of [Ni/Fe] for [Fe/H] $<$ 0, although consistent with a solar-scaled pattern, within uncertainties (see e.g., \citealt{eitner2023}); this is not seen in the GALAH sample.

Positive trends between Mn, Co and Ni and age are observed in Fig.~\ref{fig:elem_age_fe}. The production of Ni is primarily attributed to Type Ia supernovae, and a positive slope is not expected. This behavior is similar to the findings presented by \citet{Spina2021}. The same trend is present in OCs studied by \citet{CarbajoHijarrubia2024}, but not in the previous OCCASO studies. Recent 3D NLTE studies suggest that [Ni/Fe] may increase at lower metallicity due to NLTE effects \citep{Storm2025}, which could contribute to the trend in the metallicity regime covered by our sample. Manganese and Co display some mild increase with age, consistent with the delayed contribution from Type Ia SNe and the metallicity dependence of their yields. As for Mn, it is important to highlight that Galactic chemical evolution models fail to align with the NLTE abundances reported by \citet{eitner2020}, unless one assumes a threefold increase in the Mn yields from core-collapse SNe. This is clearly illustrated in the work of \citet{Palla2021} (see their Fig. 13), which also demonstrates that Mn can serve as an effective tool to differentiate between various SNIa yields, especially when precise measurements for metal-rich stars, such as those in this study, are available.

\subsection{Neutron-capture elements}
\label{sec:neutron_capture}

Elements heavier than iron are primarily formed through neutron-capture processes, with only a few proton-rich nuclei formed by photodisintegration. Two primary neutron-capture mechanisms have been identified in nature: the slow (s) and rapid (r) processes. These processes occur at opposite ends of the neutron density range, with the s process occurring at lower densities and the r process at extremely high densities. They are characterized by neutron capture timescales that are slower or faster, respectively, than the beta-decay timescale of unstable nuclei. Together, these processes account for the majority of the abundances of elements heavier than iron in the Solar System (see the recent reviews by \citealt{cowan2021} and \citealt{lugaro2023}). For the s-process, the elements are predominantly produced in AGB (via the $^{13}$C($\alpha$,$n$)$^{16}$O neutron source) or massive stars (via the $^{22}$Ne($\alpha,n$)$^{25}$Mg. For the r-process, the main production channel is still under debate (compact mergers, special CC SNe; see, e.g, \citealt{cowan2021}).

We inferred elemental abundances for the s-process elements yttrium and strontium, while barium lines were heavily saturated and susceptible to significant uncertainties. For this reason, we do not publish Ba abundances (see also discussion in \citealt{baratella2021,nordlander2024}). In our targets, along with the GALAH sample, most of the young clusters exhibit super-solar [Y/Fe] ratios. This observation, which has been previously noted in (young) open clusters \citep{Magrini2018, baratella2021}, is not unexpected. However, whether this enhancement is genuine is strongly debated in the literature \citep{Dorazi2022}. Although Sr and Y belong to the first peak of the s-process and share similar nucleosynthetic origins, they exhibit different behaviors in our analysis. In particular, Sr demonstrates a flatter distribution, with the majority of clusters showing abundances consistent with solar levels. In contrast, the GALAH open clusters show exceptionally high super-solar abundances, exceeding 0.5 dex. However, those values were inferred using the neutral Sr line at 6550\,\AA, known to be blended (see also the discussion in the GALAH DR3 paper, \citealt{Buder2021}). Such values are not consistent with any chemical evolution models. Therefore, we urge caution when interpreting these numbers, as they are subject to significant uncertainties. Europium abundances were also determined to trace the r-process contribution. Our findings indicate that these abundances are consistent with solar-scaled patterns, aligning with observations from the GALAH sample, although the GALAH data display a larger scatter. We find a slight positive slope with age, although not very significant. \citet{DelgadoMena2019} seem to indicate a trend in the opposite direction (see their Fig. 7). However, this is due to the inclusion of stars with ages larger than in our sample; when restricting to a maximum age of 4 Gyr, also their distribution is almost flat.

The interplay between the abundances of s-process elements and those of elements with opposite behaviors, such as the $\alpha$ elements, enhances their correlation with stellar age. This relationship has been highlighted, for instance, by \citet{Nissen2015, Spina2016, Spina2018, DelgadoMena2019}, who identified the [Y/Mg] ratio (among others) as a reliable chronometer for Galactic evolution. However, the empirical relationships between n-capture elements and $\alpha$ elements are not consistent across the Galactic disk. Recent studies \citep[see, e.g.][]{feltzing2017,Casali2020b, Viscasillas2022} have shown that these relationships vary depending on metallicity and the spatial location within the disk. This variation is significant, particularly given the presence of a metallicity gradient across the disk, especially in the inner part, up to 12-14 kpc (see, e.g., \citealt{Donor2020}). To accurately calibrate these empirical relations, samples with precise age determinations are essential, and open clusters represent one of the optimal choices within the high-metallicity regime (with [Fe/H] ranging from approximately -0.5 to 0.5 dex).
\begin{figure}
\centering
\includegraphics[width=\linewidth]{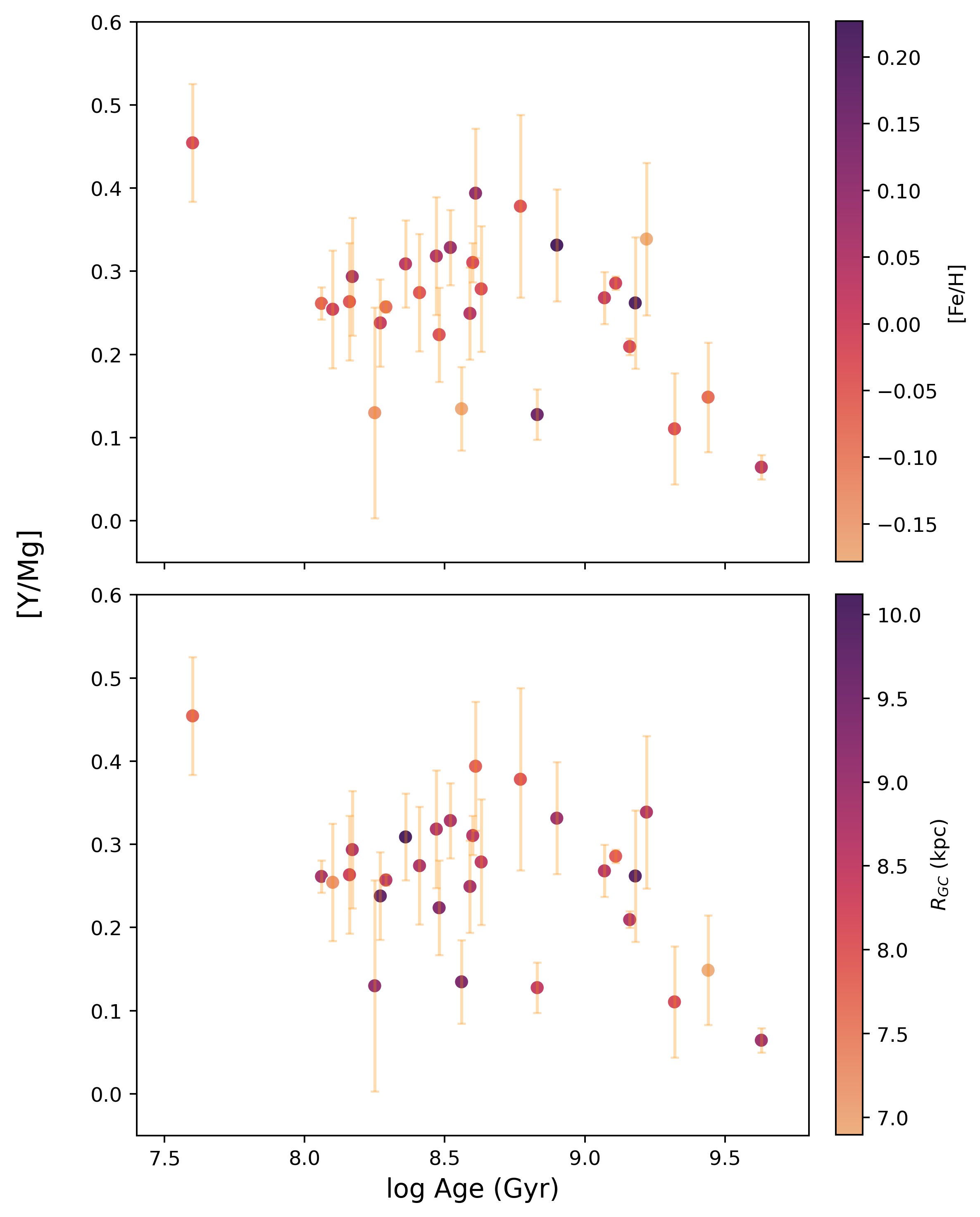}
\caption{[Y/Mg] as a function of age. The open clusters from this work are color-coded by [Fe/H] in the top panel and by $R_{\textrm{GC}}$ in the bottom panel.}
\label{fig:ymg_age}
\end{figure}
Figure \ref{fig:ymg_age} presents the [Y/Mg] ratios plotted against the ages of the open clusters analyzed in this study. 
Our analysis reveals a nearly flat trend in [Y/Mg] versus age for clusters ranging from 100 Myr to approximately 1 Gyr, irrespective of the open clusters' metallicities or their positions within the Galaxy (although we acknowledge that our data covers a limited range). The youngest cluster in our sample, Gulliver 18, with an age of about 40 Myr, displays the highest [Y/Mg] ratios, with [Y/Fe] ratios three times greater than the solar value. In contrast, M 67, the oldest cluster in our sample, exhibits the lowest [Y/Mg] ratio. Notably, the ratios for other clusters within this range remain consistently flat with age, indicating that [Y/Mg] ratios may not be effective for dating the majority of OCs.


\section{Summary } 
\label{sec:conclusions}

In this work, we examine 33 open clusters observed by the SPA project, comprising a total of 95 giant stars with high-resolution spectra from HARPS-N. We chemically characterized 13 open clusters within the SPA context and reanalyzed 20 previously studied clusters, creating a homogeneous and expanded sample within the SPA framework. We presented NLTE atmospheric parameters obtained using {\tt LOTUS} through the equivalent width method and chemical abundances obtained via spectral synthesis with {\tt TSFitPy}, covering ten elements from different nucleosynthesis groups.

We compared our results and those from the literature and found a good agreement. We explored the Galactic trends in abundance ratios [X/Fe] as a function of metallicity and age. For comparison, we used the open clusters studied by \citet{Spina2021}, but with mean abundances derived using only GALAH DR3 original data and the open clusters by \citet{Myers2022} from the OCCAM survey. In general, our results follow similar trends to those observed in GALAH and APOGEE.

Regarding the age trends, the $\alpha$ elements, such as Mg, Si, and Ti, show increasing trends with age. For the odd-Z elements, Na exhibits a decrease with age, while Al aligns more closely with the trends observed for $\alpha$ elements. The iron-peak elements, Mn, Co, and Ni, display a positive correlation with age. Neutron-capture Y and Eu shows a decreasing trend with age. These trends are consistent with expectations from GCE models including the outcomes of nucleosynthesis processes in stars, except for Ni, and agree with previous studies.

In summary, this work added nine more open clusters with a full chemical characterization, and explored the age trends in elemental abundances, positive for Mg, Si, Ti, Al, Mn, Co and Ni and negative for Na and Y. These results illustrate the important role of open clusters as valuable tracers of Galactic chemical evolution, reinforcing that NLTE corrections are critical for accurate stellar parameter and abundance determination. 

\section*{Data availability}

Tables \ref{tab:atm_params} and \ref{tab:abundances} are only available in electronic form at the CDS via anonymous ftp to \hyperlink{cdsarc.u-strasbg.fr}{cdsarc.u-strasbg.fr} (130.79.128.5) or via \hyperlink{http://cdsweb.u-strasbg.fr/cgi-bin/qcat?J/A+A/}{http://cdsweb.u-strasbg.fr/cgi-bin/qcat?J/A+A/}.

\begin{acknowledgements}
This research used the facilities of the Italian Center for Astronomical Archive (IA2) operated by INAF at the Astronomical Observatory of Trieste. This work has made use of data from the European Space Agency (ESA) mission Gaia (https://www.cosmos.esa.int/gaia), processed by the Gaia Data Processing and Analysis Consortium (DPAC, https://www.cosmos.esa.int/web/gaia/dpac/consortium). Funding for the DPAC has been provided by national institutions, in particular the institutions participating in the Gaia Multilateral
Agreement. This research has made use of the VizieR catalog access tool, CDS, Strasbourg, France (DOI : 10.26093/cds/vizier). The original description of the VizieR service was published in 2000, A\&AS 143, 23. Use of the NASA's Astrophysical Data System and TOPCAT \citep{topcat} are also acknowledged. We acknowledge funding from INAF MiniGrant 2022 (High resolution spectroscopy of open clusters) and the INAF grant Open Clusters and stellar structures in the local Galactic disk (ref. Antonella Vallenari) (CRA 1.05.23.05.19).
\end{acknowledgements}

\bibliographystyle{aa}
\bibliography{refs}

\begin{appendix}

\onecolumn

\section{Properties of the individual stars}

\begin{table*}[ht]
\centering
\caption{Properties of the stars of the new open clusters within the SPA project. The table contains 
information from $Gaia$ DR3, exposure times in seconds, resulting S/N, and RV from iSpec.}
\label{tab:sample_prop}
\setstretch{1.15}
\scalebox{0.83}{
\begin{tabular}{lrrrrrrrrrr}
\hline \hline
\multicolumn{1}{c}{Name} & \multicolumn{1}{c}{Gaia DR3 ID} &  \multicolumn{1}{c}{RA} &  \multicolumn{1}{c}{DEC} & \multicolumn{1}{c}{G} & \multicolumn{1}{c}{G$_{BP}$} & \multicolumn{1}{c}{G$_{RP}$} & \multicolumn{1}{c}{Exp.T} & \multicolumn{1}{c}{S/N} & \multicolumn{1}{c}{RV$_{Gaia}$} & \multicolumn{1}{c}{RV$_{spec}$} \\
\hline
\multicolumn{11}{c}{\textbf{IC 4756}} \\
\#0  & 4283940671842998272 & 279.682 & 5.239  & 9.05  & 9.60  & 8.35 & 3600 & 92 & -25.29 $\pm$ 0.15 & -25.66 $\pm$ 0.02 \\
\#2  & 4283984931511649792 & 279.399 & 5.260  & 7.52  & 8.25  & 6.69 & 690 & 148  & -25.08 $\pm$ 0.12 & -25.10 $\pm$ 0.01 \\
\#4  & 4283901746580386816 & 279.138 & 5.212  & 8.49  & 9.23  & 7.65 & 2100 & 125 & -24.98 $\pm$ 0.13 & -25.04 $\pm$ 0.01 \\
\#5  & 4283961979205354496 & 280.077 & 5.314  & 8.90  & 9.48  & 8.18 & 2100 & 174 & -24.98 $\pm$ 0.15 & -25.31 $\pm$ 0.01 \\
\#7  & 4284658038773033728 & 278.948 & 5.338  & 9.10  & 9.73  & 8.33 & 1380 & 121 & -25.00 $\pm$ 0.22 & -25.15 $\pm$ 0.01 \\
\#8  & 4283900062953084800 & 279.158 & 5.136  & 9.10  & 9.73  & 8.33 & 1380 & 133 & -27.75 $\pm$ 0.34 & -27.23 $\pm$ 0.01 \\
\#9  & 4284802074805175936 & 279.621 & 5.981  & 9.13  & 9.67  & 8.44 & 1400 & 90 & -25.21 $\pm$ 0.22 & -25.61 $\pm$ 0.02 \\
\#11 & 4283789905627894272 & 278.891 & 4.450  & 9.42  & 10.04 & 8.66 & 1800 & 92 & -25.07 $\pm$ 0.32 & -25.13 $\pm$ 0.02 \\
\#12 & 4283983552796484864 & 279.376 & 5.204  & 9.44  & 10.02 & 8.70 & 1800 & 86 & --                & -25.88 $\pm$ 0.02 \\
\#13 & 4284806438475643776 & 279.337 & 5.895  & 9.17  & 9.67  & 8.44 & 1800 & 115 & -24.50 $\pm$ 0.14 & -24.75 $\pm$ 0.02 \\
\#14 & 4283997575895463680 & 279.272 & 5.292  & 9.41  & 10.00 & 8.67 & 1800 & 115 & -25.18 $\pm$ 0.16 & -25.73 $\pm$ 0.01 \\
\#15 & 4286303114338968448 & 279.851 & 6.122  & 8.71  & 9.28  & 7.98 & 1200 & 110 & -25.56 $\pm$ 0.13 & -25.50 $\pm$ 0.01 \\
\hline
\multicolumn{11}{c}{\textbf{NGC 2632}} \\
\#1 & 661271173693364864   & 129.961 & 19.540 & 6.39  & 6.82  & 5.72 & 600 & 188 & 35.44 $\pm$ 0.14 & 35.31 $\pm$ 0.01 \\
HD 73665 & 661324431287688448 & 130.026 & 20.007 & 6.15 & 6.62 & 5.52 & 1800 & 144 & 33.92 $\pm$ 0.13 & 33.97 $\pm$ 0.01 \\
\hline
\multicolumn{11}{c}{\textbf{NGC 6800}} \\
\#1  & 2023140603917404416 & 291.570 & 25.157 & 10.27 & 10.94 & 9.48 & 3600 & 71 & -14.15 $\pm$ 0.21 & -14.34 $\pm$ 0.02 \\
\hline
\multicolumn{11}{c}{\textbf{NGC 6991}} \\
\#4  & 2166845227448342400 & 313.989 & 47.395 & 9.39  & 9.90  & 8.73 & 2800 & 95 & -11.68 $\pm$ 0.21 & -12.02 $\pm$ 0.01 \\
\#6  & 2167008230043576192 & 313.951 & 47.910 & 9.54  & 10.08 & 8.85 & 2800 & 85 & -12.64 $\pm$ 0.15 & -12.76 $\pm$ 0.01 \\
\#7  & 2166853228954093312 & 313.542 & 47.382 & 9.60  & 10.11 & 8.92 & 2800 & 86 & --                & -12.42 $\pm$ 0.01 \\
\#9  & 2166908105771609472 & 313.348 & 47.670 & 9.76  & 10.30 & 9.06 & 2800 & 75 & -12.87 $\pm$ 0.13 & -12.95 $\pm$ 0.01 \\
\#10 & 2166821789793563904 & 313.462 & 47.085 & 9.81  & 10.38 & 9.10 & 2800 & 87 & -12.44 $\pm$ 0.18 & -12.65 $\pm$ 0.01 \\
\hline
\multicolumn{11}{c}{\textbf{NGC 7086}} \\
\#1  & 2171663253032105984 & 322.486 & 51.223 & 9.73  & 10.65 & 8.80 & 1400 & 79 & -22.06 $\pm$ 0.17 & -22.09 $\pm$ 0.02 \\
\#2  & 2171707542735344000 & 322.589 & 51.610 & 9.44  & 10.54 & 8.41 & 2800 & 89 & -20.60 $\pm$ 0.13 & -20.32 $\pm$ 0.02 \\
\hline
\multicolumn{11}{c}{\textbf{NGC 752}} \\
\#3  & 342937195667536512  & 29.263  & 38.134 & 8.65  & 9.15  & 7.99 & 2100 & 228 & 5.76 $\pm$ 0.15 & 5.38 $\pm$ 0.01 \\
\#4  & 342899537393760512  & 29.812  & 38.015 & 8.73  & 9.23  & 8.07 & 2100 & 194 & 4.86 $\pm$ 0.15 & 4.75 $\pm$ 0.01 \\
\#5  & 342890127122193280  & 29.720  & 37.816 & 8.78  & 9.30  & 8.11 & 2100 & 230 & 5.93 $\pm$ 0.16 & 5.86 $\pm$ 0.01 \\
\#6  & 344569012659424512  & 31.304  & 39.396 & 9.17  & 9.62  & 8.55 & 1500 & 160 & -4.90 $\pm$ 0.20 & -5.03 $\pm$ 0.02 \\
\#7  & 325299727783639552  & 35.419  & 33.476 & 9.29  & 9.85  & 8.58 & 2100 & 162 & 52.46 $\pm$ 0.18 & 52.15 $\pm$ 0.01 \\
\hline
\multicolumn{11}{c}{\textbf{UBC 3}} \\
\#1 & 4505874693061489280 & 283.793 & 12.329 & 10.02 & 10.82 & 9.14 & 3600 & 83 & 1.92 $\pm$ 0.18 & 2.09 $\pm 0.02$ \\
\hline
\multicolumn{11}{c}{\textbf{UBC 60}} \\
\#3  & 179623851673246464  & 67.885  & 39.370 & 10.11 & 10.87 & 9.26 & 3000 & 98 & -6.18  $\pm$ 0.17 & -6.08  $\pm$ 0.02 \\
\hline
\multicolumn{11}{c}{\textbf{UBC 131}} \\
\#1  & 1814336194629117824 & 310.234 & 20.027 & 10.25 & 10.75 & 9.59  & 4200 & 90 & 23.89 $\pm$ 0.19 & 23.88 $\pm$ 0.01 \\
\#2  & 1817359031271791232 & 310.261 & 20.482 & 10.66 & 11.16 & 10.01 & 4200 & 73 & 23.66 $\pm$ 0.20 & 23.71 $\pm$ 0.01 \\
\#3  & 1814243324558277248 & 310.049 & 19.942 & 10.71 & 11.20 & 10.06 & 4200 & 63 & 24.32 $\pm$ 0.22 & 24.19 $\pm$ 0.02 \\
\hline
\multicolumn{11}{c}{\textbf{UBC 141}} \\
\#1  & 1869885794120648192 & 313.475 & 35.695 & 11.51 & 12.22 & 10.68 & 4830 & 43 & -18.43 $\pm$ 0.23 & -18.59 $\pm$ 0.01 \\
\hline
\multicolumn{11}{c}{\textbf{UBC 169}} \\
\#1  & 1988520659206227840 & 342.615 & 49.468 & 9.75  & 10.33 & 9.03 & 2800 & 76 & -30.02 $\pm$ 0.15 & -30.13 $\pm$ 0.03 \\
\#2  & 1988520899724392064 & 342.671 & 49.518 & 9.74  & 10.31 & 9.03 & 2800 & 74 & -29.74 $\pm$ 0.15 & -29.85 $\pm$ 0.04 \\
\hline
\multicolumn{11}{c}{\textbf{UBC 170}} \\
\#1  & 2007935075311695872 & 337.878 & 58.021 & 10.16 & 10.98 & 9.27 & 4200 & 86 & -28.40 $\pm$ 0.16 & -28.59 $\pm$ 0.03 \\
\#2  & 2007981499630067456 & 337.770 & 58.042 & 10.42 & 11.23 & 9.53 & 4200 & 72 & -28.50 $\pm$ 0.24 & -28.41 $\pm$ 0.03 \\
\#3  & 2007934044519547904 & 337.712 & 57.992 & 10.42 & 11.21 & 9.55 & 2846 & 78 & -28.51 $\pm$ 0.17 & -28.84 $\pm$ 0.02 \\
\hline
\multicolumn{11}{c}{\textbf{UBC 194}} \\
\#1  & 354343808469156608  & 37.709  & 48.284 & 9.63  & 10.22 & 8.89 & 3600 & 131 & -29.85 $\pm$ 0.18 & -30.12 $\pm$ 0.04 \\
\hline
\multicolumn{11}{c}{\textbf{UBC 577}} \\
\#1  & 4531532690223665664 & 282.260 & 22.216 & 11.07 & 11.64 & 10.34 & 4830 & 35 & -9.04 $\pm$ 0.17 & -9.33 $\pm$ 0.01 \\
\#2  & 4531526058785223424 & 282.158 & 22.124 & 11.28 & 11.87 & 10.55 & 4830 & 58 & -8.90 $\pm$ 0.19 & -9.06 $\pm$ 0.01 \\
\#3  & 4531525337240008576 & 282.118 & 22.059 & 10.36 & 10.94 & 9.63  & 5400 & 74 & -9.24 $\pm$ 0.14 & -9.55 $\pm$ 0.02 \\
\#5  & 4531474038155634560 & 281.974 & 21.654 & 11.02 & 11.57 & 10.31 & 4200 & 45 & -9.26 $\pm$ 0.19 & -9.34 $\pm$ 0.01 \\
\hline
\end{tabular}
}
\end{table*}

\section{LOTUS Markov-Chain Monte Carlo corner plot}

\begin{figure}[ht]
\centering
\includegraphics[width=0.65\linewidth]{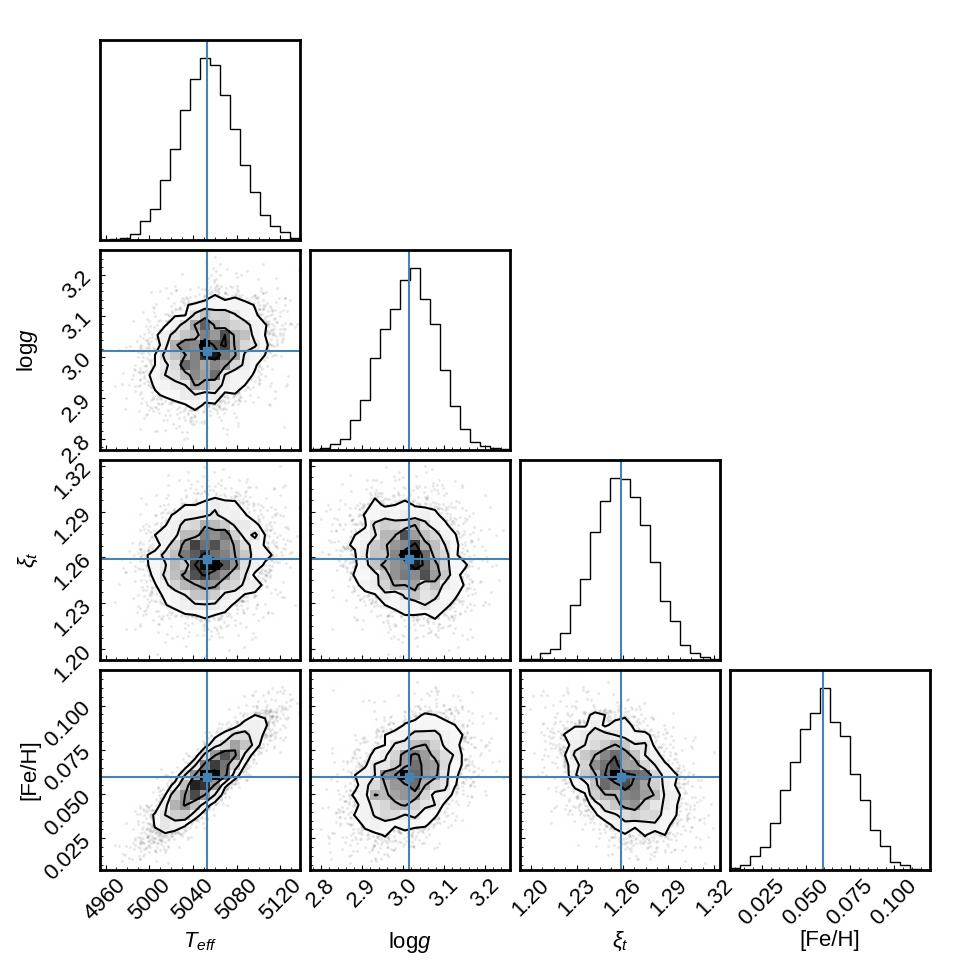}
\caption{Marginalized posterior distributions of the stellar parameters for IC 4756 star \#5, with the blue intersecting lines indicating the optimized parameters obtain by {\tt LOTUS}.}
\label{fig:mcmc_lotus}
\end{figure}

\section{Comparison of LTE and NLTE atmospheric parameters}

\begin{figure}[ht]
\centering
\includegraphics[width=0.8\linewidth]{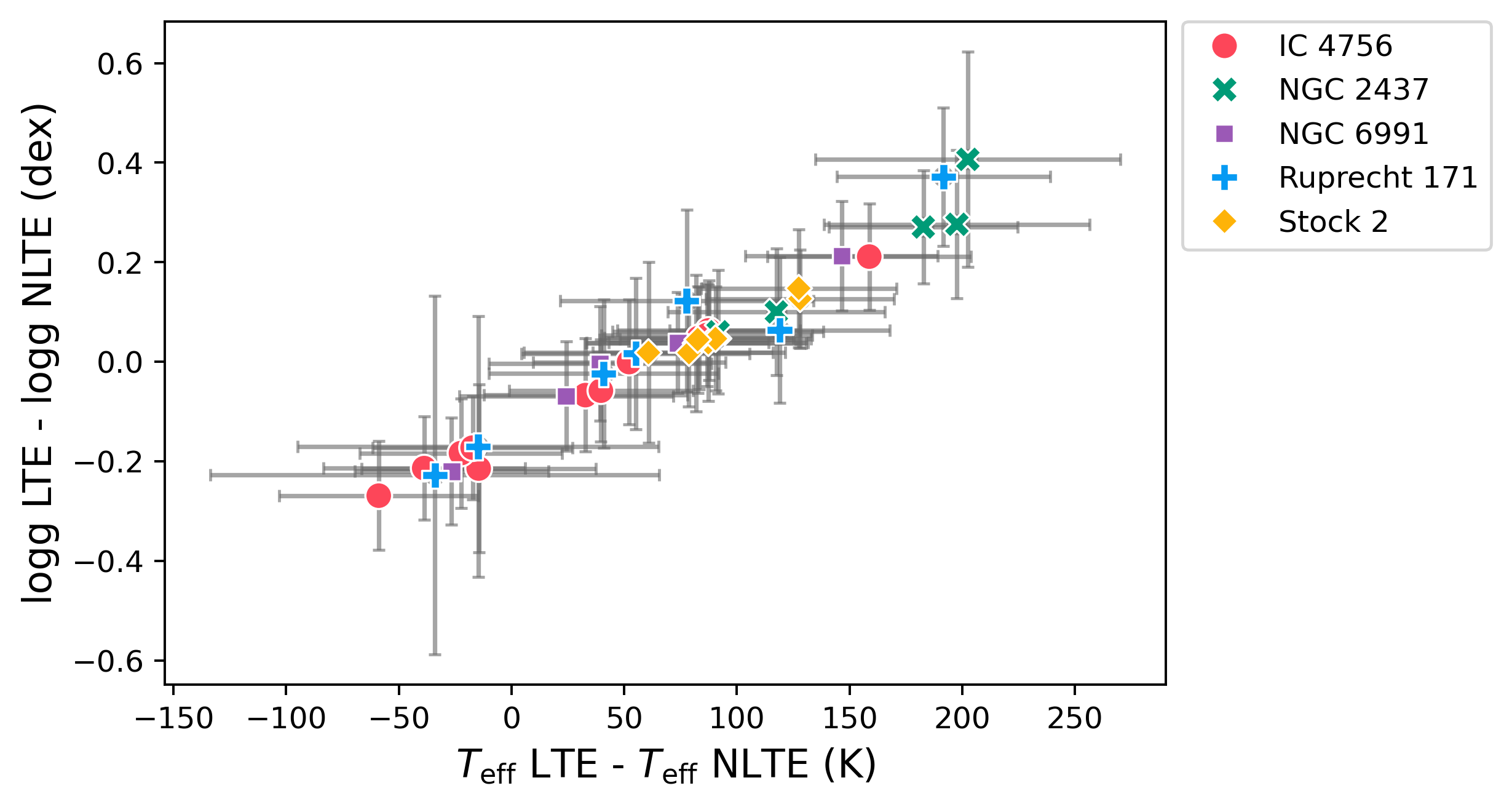}
\caption{Differences between LTE and NLTE atmospheric parameters obtained with {\tt LOTUS} for stars in five open clusters.}
\label{fig:lte_nlte_params}
\end{figure}

\newpage

\section{Line list}

\begin{table}[ht]
\centering
\caption{Line list for the {\tt TSFitPy} synthesis.}
\begin{tabular}{cccr|cccr}
\hline \hline
\multicolumn{1}{c}{Wavelength ($\AA$)} & \multicolumn{1}{c}{Ion} & \multicolumn{1}{c}{E.P. (eV)} & \multicolumn{1}{c}{log $gf$} 
&\multicolumn{1}{c}{Wavelength ($\AA$)} & \multicolumn{1}{c}{Ion} & \multicolumn{1}{c}{E.P. (eV)} & \multicolumn{1}{c}{log $gf$} 
\\
\hline
5688.200 & 11.01 & 2.104 & -0.404 & 6554.223 & 22.01 & 1.443 & -1.150\\
6154.226 & 11.01 & 2.102 & -1.547 & 4911.193 & 22.02 & 3.124 & -0.640\\
6160.747 & 11.01 & 2.104 & -1.246 & 5418.751 & 22.02 & 1.582 & -2.130\\
5711.088 & 12.01 & 4.346 & -1.742 & 4761.506 & 25.01 & 2.953 & -0.548\\
6696.023 & 13.01 & 3.143 & -1.569 & 4765.852 & 25.01 & 2.941 & -0.445\\
6698.673 & 13.01 & 3.143 & -1.870 & 4813.476 & 27.01 & 3.216 & -0.191\\
5675.417 & 14.01 & 5.619 & -1.234 & 4867.869 & 27.01 & 3.117 & -0.193\\
5684.484 & 14.01 & 4.954 & -1.607 & 5369.588 & 27.01 & 4.259 & -1.174\\
5701.104 & 14.01 & 4.930 & -1.981 & 5483.353 & 27.01 & 1.710 & -2.234\\
5772.146 & 14.01 & 5.082 & -1.643 & 4829.023 & 28.01 & 3.542 & -0.140\\
5793.073 & 14.01 & 4.930 & -1.894 & 4913.973 & 28.01 & 3.743 & -0.500\\
5948.541 & 14.01 & 5.082 & -1.230 & 4935.830 & 28.01 & 3.941 & -0.213\\
6555.462 & 14.01 & 5.984 & -0.886 & 5748.351 & 28.01 & 1.676 & -3.240\\
6721.848 & 14.01 & 5.863 & -0.940 & 5805.217 & 28.01 & 4.167 & -0.579\\
4758.118 & 22.01 & 2.249 &  0.510 & 6007.310 & 28.01 & 1.676 & -3.400\\
4913.614 & 22.01 & 1.873 &  0.220 & 6108.116 & 28.01 & 1.676 & -2.600\\
4999.503 & 22.01 & 0.826 &  0.320 & 6128.973 & 28.01 & 1.676 & -3.430\\
5016.161 & 22.01 & 0.848 & -0.480 & 6175.366 & 28.01 & 4.089 & -0.389\\
5024.844 & 22.01 & 0.818 & -0.530 & 6482.798 & 28.01 & 1.935 & -2.630\\
5866.451 & 22.01 & 1.067 & -0.790 & 6586.310 & 28.01 & 1.951 & -2.780\\
5953.160 & 22.01 & 1.887 & -0.273 & 4607.330 & 38.01 & 0.000 &  0.283\\
5965.828 & 22.01 & 1.879 & -0.353 & 4883.682 & 39.02 & 1.084 &  0.190\\
5978.541 & 22.01 & 1.873 & -0.440 & 4900.119 & 39.02 & 1.033 &  0.030\\
6261.098 & 22.01 & 1.430 & -0.530 & 6645.064 & 63.02 & 1.380 & -1.814\\
\hline
\end{tabular}
\label{tab:linelist_tsfit}
\end{table}

\newpage

\section{Abundance sensitivities}

\begin{table}[ht]
\centering
\caption{Abundance sensitivities to change in stellar parameters for a sub-sample of stars.}
\begin{tabular}{clrrrr}
\hline \hline
Species & \multicolumn{1}{c}{Star} & \multicolumn{1}{c}{$\Delta$ $T_{\rm eff}$} & \multicolumn{1}{c}{$\Delta$ log g} & \multicolumn{1}{c}{$\Delta$ [Fe/H]} & \multicolumn{1}{c}{$\Delta$ $\xi$}  \\
& & (+100 K) & (+0.2 dex) & (+0.2 dex) & (+0.2 km $\rm s^{-1}$) \\
\hline 
\multirow{5}{*}{[Mg/Fe]} & IC 4756 \#4 & 0.063 & 0.005 & -0.201 & -0.094 \\
& IC 4756 \#5 & 0.062 & -0.021 & -0.201 & -0.073 \\
& IC 4756 \#9 & 0.065 & -0.022 & -0.196 & -0.064 \\
& NGC 752 \#3 & 0.070 & -0.012 & -0.197 & -0.076 \\
& NGC 6800 \#1 & 0.068 & -0.006 & -0.204 & -0.073 \\
\hline
\multirow{5}{*}{[Si/Fe]} & IC 4756 \#4 & -0.007 & 0.071 & -0.254 & -0.037 \\
& IC 4756 \#5 & 0.018 & 0.043 & -0.212 & -0.046 \\
& IC 4756 \#9 & 0.029 & 0.062 & -0.192 & -0.015 \\
& NGC 752 \#3 & -0.039 & 0.008 & -0.271 & -0.095 \\
& NGC 6800 \#1 & 0.003 & 0.050 & -0.221 & -0.044 \\
\hline
\multirow{5}{*}{[Ti/Fe]} & IC 4756 \#4 & 0.056 & -0.015 & -0.025 & -0.187 \\
& IC 4756 \#5 & 0.043 & -0.005 & -0.234 & -0.183 \\
& IC 4756 \#9 & 0.044 & -0.007 & -0.373 & -0.127 \\
& NGC 752 \#3 & 0.071 & -0.041 & -0.403 & -0.142 \\
& NGC 6800 \#1 & 0.044 & -0.028 & -0.308 & -0.111 \\
\hline
\multirow{5}{*}{[Na/Fe]} & IC 4756 \#4 & 0.083 & -0.011 & -0.210 & -0.069 \\
& IC 4756 \#5 & 0.071 & -0.019 & -0.211 & -0.049 \\
& IC 4756 \#9 & 0.069 & -0.018 & -0.205 & -0.042 \\
& NGC 752 \#3 & 0.078 & -0.015 & -0.211 & -0.050 \\
& NGC 6800 \#1 & 0.074 & -0.013 & -0.220 & -0.057 \\
\hline
\multirow{5}{*}{[Al/Fe]} & IC 4756 \#4 & 0.074 & -0.007 & -0.205 & -0.033 \\
& IC 4756 \#5 & 0.061 & -0.007 & -0.211 & -0.018 \\
& IC 4756 \#9 & 0.060 & -0.005 & -0.199 & -0.015 \\
& NGC 752 \#3 & 0.068 & -0.003 & -0.222 & -0.019 \\
& NGC 6800 \#1 & 0.064 & -0.001 & -0.237 & -0.019 \\
\hline
\multirow{5}{*}{[Mn/Fe]} & IC 4756 \#4 & 0.102 & -0.004 & -0.194 & -0.163 \\
& IC 4756 \#5 & 0.080 & -0.016 & -0.213 & -0.125 \\
& IC 4756 \#9 & 0.089 & -0.009 & -0.207 & -0.101 \\
& NGC 752 \#3 & 0.094 & -0.028 & -0.212 & -0.143 \\
& NGC 6800 \#1 & 0.091 & -0.024 & -0.213 & -0.129 \\
\hline
\multirow{5}{*}{[Co/Fe]} & IC 4756 \#4 & 0.039 & 0.028 & -0.176 & -0.079 \\
& IC 4756 \#5 & 0.078 & 0.018 & -0.190 & -0.040 \\
& IC 4756 \#9 & 0.093 & 0.019 & -0.116 & -0.021 \\
& NGC 752 \#3 & 0.069 & 0.000 & -0.185 & -0.053 \\
& NGC 6800 \#1 & 0.079 & 0.002 & -0.189 & -0.047 \\
\hline
\multirow{5}{*}{[Ni/Fe]} & IC 4756 \#4 & 0.089 & 0.034 & -0.212 & -0.122 \\
& IC 4756 \#5 & 0.111 & -0.004 & -0.220 & -0.103 \\
& IC 4756 \#9 & 0.110 & -0.014 & -0.211 & -0.09 \\
& NGC 752 \#3 & 0.114 & 0.001 & -0.223 & -0.112 \\
& NGC 6800 \#1 & 0.120 & -0.018 & -0.233 & -0.105 \\
\hline
\multirow{5}{*}{[Sr/Fe]} & IC 4756 \#4 & -0.107 & -0.258 & 0.009 & -0.439 \\
& IC 4756 \#5 & -0.044 & -0.196 & -0.442 & -0.290 \\
& IC 4756 \#9 & -0.030 & -0.162 & -0.412 & -0.244 \\
& NGC 752 \#3 & -0.066 & -0.236 & -0.469 & -0.335 \\
& NGC 6800 \#1 & -0.066 & -0.227 & -0.451 & -0.299 \\
\hline
\multirow{5}{*}{[Y/Fe]} & IC 4756 \#4 & 0.050 & 0.087 & -0.196 & -0.168 \\
& IC 4756 \#5 & 0.060 & 0.089 & -0.129 & -0.109 \\
& IC 4756 \#9 & 0.012 & 0.074 & -0.152 & -0.148 \\
& NGC 752 \#3 & 0.044 & 0.076 & -0.143 & -0.139 \\
& NGC 6800 \#1 & -0.022 & 0.074 & -0.208 & -0.180 \\
\hline
\multirow{5}{*}{[Eu/Fe]} & 
IC 4756 \#4 & -0.010    & 0.110 &  -0.160   & -0.020  \\
& IC 4756 \#5 & -0.010  & 0.090 &  -0.140   &  -0.010\\
& IC 4756 \#9 & -0.010  & 0.080 &  -0.150  &  -0.010\\
& NGC 752 \#3 &  -0.010 & 0.080 &   -0.140 &  -0.010 \\
& NGC 6800 \#1 &  -0.010& 0.090 &   -0.120  &  -0.020\\
\hline
\end{tabular}
\label{tab:sens}
\end{table}

\newpage

\section{Abundance trends}

\begin{figure*}[htbp!]
\centering
\includegraphics[width=0.5\textwidth]{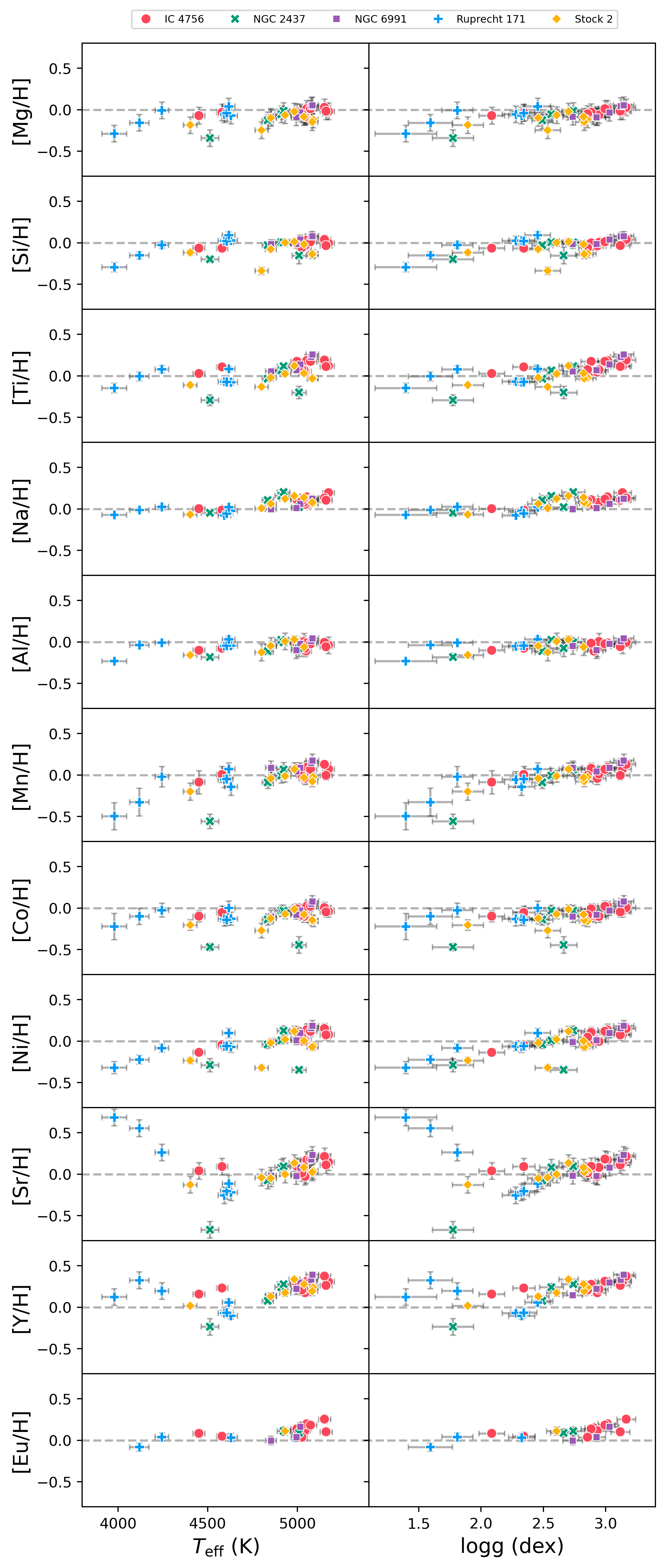}
\caption{Abundances [X/H] as a function of $T_{\rm eff}$ (first column) and log $g$ (second column). Each symbol represent a different open cluster. }
\label{fig:abund_trends}
\end{figure*}

\newpage

\section{Comparison with the literature}

This table summarizes the comparison with the literature. For $T_{\rm eff}$ and log $g$ we computed the mean differences (this work - literature) for each OC using the stars in common (N) and the error is the standard deviation. For metallicity and element abundances, we determined the differences by computing the mean cluster abundances derived in this work minus the same from the literature.

\begin{table*}[ht]
\caption{Differences of the atmospheric parameters and abundances for the OCs in common with the literature.}
\label{tab:comparison_literature}
\setstretch{1.22}
\scalebox{0.56}{
\begin{tabular}{lrrcrrrrrrrrrrrrl}
\hline \hline
OC & $T_{\rm eff}$ (K) & $\log g$ (dex) & N & $\rm [Fe/H]$ & $\rm [Mg/Fe]$ & $\rm [Si/Fe]$ & $\rm [Ti/Fe]$ & $\rm [Na/Fe]$ & $\rm [Al/Fe]$ & $\rm [Mn/Fe]$ & $\rm [Co/Fe]$ & $\rm [Ni/Fe]$ & $\rm [Sr/Fe]$ & $\rm [Y/Fe]$ & $\rm [Eu/Fe]$ & Reference  \\
\hline
\multicolumn{17}{c}{\Large \textbf{New in SPA}} \\
\hline
\multirow{9}{*}{IC 4756} & 8 $\pm$ 84 & 0.58 $\pm$ 0.24 & 2 & 0.15 & - & -0.36 & - & -0.48 & -0.34 & - & - & -0.03 & - & - & - & \citet{Jacobson2007} \\
& -69 $\pm$ 5 & 0.00 $\pm$ 0.01 & 2 & -0.03 & - & - & - & - & - & - & - & - & - & - & - & \citet{Santos2009} \\
& - & - & - & -0.04 & 0.02 & -0.08 & 0.15 & 0.10 & - & - & -0.11 & 0.06 & - & - & - & \citet{Smiljanic2009} \\
& -65 $\pm$ 12 & 0.01 $\pm$ 0.08 & 2 & -0.08 & - & -0.12 & 0.00 & -0.10 & -0.02 & - & - & 0.01 & - & - & - & \citet{Pace2010} \\
& -138 $\pm$ 90 & -0.49 $\pm$ 0.34 & 6 & 0.01 & -0.15 & -0.15 & - & -0.12 & -0.17 & - & - & 0.02 & - & - & - & \citet{Ting2012} \\
& -48 $\pm$ 46 & 0.27 $\pm$ 0.11 & 8 & 0.02 & -0.11 & -0.06 & 0.10 & -0.05 & -0.09 & 0.10 & -0.01 & 0.11  & - & 0.22 & 0.05 & \citet{Bagdonas2018} \\
& -71 $\pm$ 87 & -0.09 $\pm$ 0.20 & 4 & -0.00 & 0.01 & -0.02 & 0.07 & - & - & - & - & 0.08 & - & - & - & \citet{Casamiquela2017, Casamiquela2019} \\
& - & - & - & 0.05 & -0.05 & -0.08 & - & - & 0.10 & - & - & - & - & - & - & \citet{Ray2022}\\
& 50 $\pm$ 49 & 0.21 $\pm$ 0.11 & 8 & -  & -  & - & - & - & - & 0.10  & - & - & - & - & - & \citet{Tsantaki2023} \\
& -43 $\pm$ 35 & 0.05 $\pm$ 0.09 & 4 & 0.03 & 0.02 & -0.07 & 0.11 & -0.06 & -0.09 & 0.17 & -0.01 & 0.09 & -0.15 & 0.20 & - & \citet{CarbajoHijarrubia2024} \\
\hline
\multirow{3}{*}{NGC 6991} & -  & - & - & 0.03 & 0.01 & -0.04 & 0.15 & -0.07 & -0.14 & 0.08 & -0.06 & 0.07 & - & 0.15 & -0.07 & \citet{Reddy2019} \\
& 10 & 0.27 & 1 & 0.03 & -0.04 & -0.02 & 0.07 & - & - & - & - & 0.09 & -  & - & - & \citet{Casamiquela2017, Casamiquela2019} \\
& -  & - & - & 0.07 & -0.02 & -0.06 & 0.10 & -0.11 & -0.12 & 0.15 & -0.03 & 0.06 & -0.16 & 0.16 & - & \citet{CarbajoHijarrubia2024} \\
\hline
\multirow{10}{*}{NGC 752} & - & - & - & 0.01  & - & - & - & - & - & - & - & - & - & - & - & \citet{Sestito2004} \\
& 87 & -0.29 & 1 & -0.06 & -0.17 & -0.01 & 0.14 & -0.01 & 0.05 & - & -0.05 & 0.06 & - & 0.24 & - & \citet{Carrera2011} \\
& 59 $\pm$ 31  & 0.36 $\pm$ 0.08 & 2 & 0.06 & -0.04 & -0.10 & 0.16 & -0.12 & -0.22 & 0.18 & -0.02 & 0.06 & - & 0.18 & 0.01 &\citet{Reddy2012} \\
& 67 $\pm$ 47 & 0.36 $\pm$ 0.12 & 3 & 0.01 & 0.04 & -0.10 & 0.15 & -0.05 & -0.10 & 0.28 & 0.08  & 0.03 & - & 0.27 & 0.03 & \citet{Bocek2015} \\
& 29 $\pm$ 10 & 0.13 $\pm$ 0.01 & 2 & 0.01 & -0.03 & -0.01 & 0.07 & - & - & - & - & 0.03 & - & - & - & \citet{Casamiquela2017, Casamiquela2019} \\
& -91 $\pm$ 31 & 0.31 $\pm$ 0.06 & 2 & 0.03 & -0.09 & -0.06 & 0.01 & 0.09  & -0.27 & 0.06 & -0.15 & 0.07 & - & 0.19 & -& \citet{Lum2019} \\
& - & - & - & 0.06 & 0.04  & 0.04 & 0.17 & -0.05 & -0.11 & 0.05 & 0.10 & 0.05 & - & - & - & \citet{Spina2021} \\
& - & - & - & 0.08 & -0.05 & 0.02 & 0.15 & -0.11 & -0.09 & 0.01 & 0.02 & 0.08 & - & - & - & \citet{Myers2022} \\
& - & - & - & 0.06 & 0.04  & 0.03 & 0.17 & -0.04 & -0.02 & 0.03 & 0.07 & 0.08 & - & - & - & \citet{Donor2020} \\
& -8 $\pm$ 11 & 0.17 $\pm$ 0.05 & 2 & 0.04 & -0.01 & -0.05 & 0.09 & -0.10 & -0.14 & 0.14 & -0.01 & 0.08 & -0.09 & 0.18 & - & \citet{CarbajoHijarrubia2024} \\
\hline
UBC 3 & -73 & 0.23 & 1 & 0.01 & -0.11 & -0.18 & 0.02 & -0.15 & -0.09 & 0.04 & -0.10 & -0.00 & -0.26 & 0.11 & - & \citet{CarbajoHijarrubia2024} \\
\hline
\multicolumn{17}{c}{\Large \textbf{SPA previous studies}} \\
\hline
\multirow{2}{*}{Collinder 350} & 55 $\pm$ 59 & 0.26 $\pm$ 0.06 & \multirow{2}{*}{2} & 0.00 & -0.13 & -0.08 & -0.05 & -0.16 & 0.03 & - & -0.07 & -0.07 & - & 0.25 & 0.23 & \multirow{4}{*}{\citet{Casali2020a}} \\
& 10 $\pm$ 174 & 0.22 $\pm$ 0.21 & & -0.03 & & & & & & & & & & \\
Gulliver 51 & 156 & -0.31 & \multirow{2}{*}{1} & -0.09 & -0.10 & 0.04 & 0.05 & -0.13 & -0.11 & - & -0.06 & 0.00 & - & 0.02 & 0.03 &  \\
& 66 & -0.52 & & -0.014 & & & & & & & & & & \\
NGC 7044 & -34$\pm$ 83 & 0.14 $\pm$ 0.14 & \multirow{2}{*}{4} & - & - & - & - & - & - & - & - & - & - & - & - & \\
& 28 $\pm$ 73 & -0.25 $\pm$ 0.09 & & -0.05 & & & & & & & & & & & \\
Ruprecht 171 & -136 $\pm$ 131 & -0.22 $\pm$ 0.32 & \multirow{2}{*}{7} & -0.13 & -0.03 & 0.00 & 0.02 & -0.09 & 0.00 & - & -0.13 & -0.05 & - & 0.13 & 0.10 &  \\
& -25 $\pm$ 113 & -0.24 $\pm$ 0.14 & & -0.10 & & & & & & & & & & \\
\hline
Alessi 1 & -166 $\pm$ 125 & -0.23 $\pm$ 0.31 & 4 & -0.01 & 0.11 & -0.14 & 0.15 & 0.12 & 0.11 & - & - & - & - & - & - & \multirow{16}{*}{\citet{Zhang2021,Zhang2022}} \\ 
Alessi Teutsch 11 &	-43	& 0.051 & 1 & 0.15 & 0.07 & -0.25 & 0.15 & -0.12 & 0.02  & - & - & - & - & - & - & \\ 
Basel 11 b & -211 $\pm$ 20 & -0.45 $\pm$ 0.18 & 3 & 0.04 & 0.08 & -0.17 & 0.05 & 0.00 & 0.07  & - & - & - & - & - & - & \\   
COIN-Gaia 30 & -404 & -0.99 & 1 & -0.07 & 0.11  & -0.09 & -0.08 & -0.04 & 0.06  & - & - & - & - & - & - & \\ 
Collinder 350 & -60	$\pm$ 80 & 0.13 $\pm$ 0.31 & 2 & 0.16 & -0.10 & -0.10 & -0.05 & -0.07 & 0.01  & - & - & - & - & - & - & \\   
Collinder 463 & -134 $\pm$ 54 & -0.20 $\pm$ 0.13 & 2 & 0.09 & 0.07 & -0.18 & 0.11  & -0.05 & 0.06  & - & - & - & - & - & - & \\    
Gulliver 18	& -401 & -0.98 & 1 & 0.09 & 0.22 & -0.21 & -0.16 & -0.05 & -0.14 & - & - & - & - & - & - & \\    
Gulliver 24	& -191 & -0.73 & 1 & 0.04 & 0.19 & -0.29 & 0.02 & 0.08 & -0.02 & - & - & - & - & - & - & \\   
NGC 2437 & -184 $\pm$ 80 & -0.34 $\pm$ 0.17 & 5 & -0.03 & 0.10 & -0.16 & 0.07 & 0.06 & -0.00 & - & - & - & - & - & - & \\   
NGC 2509 & 68 & 0.35 & 1 & 0.31 & 0.06 & -0.17 & 0.18 & -0.09 & -0.08 & - & - & - & - & - & - & \\ 
NGC 2548 & -156 $\pm$ 149 & -0.22 $\pm$ 0.46 & 3 & 0.07 & 0.10 & -0.06 & 0.06 & 0.08 & 0.07  & - & - & - & - & - & - & \\   
NGC 2682 & -71 $\pm$ 157 & 0.03 $\pm$ 0.30 & 2 & 0.05 & 0.05 & -0.06 & 0.19 & 0.04 & 0.05 & - & - & - & - & - &  - & \\    
NGC 7209 & -168 & 0.37 & 1 & 0.01 & 0.07 & -0.15 & 0.04 & 0.17 & -0.00 & - & - & - & - & - & - & \\
Tombaugh 5 & -197 $\pm$ 223 & -0.44 0.56 $\pm$ & 3 & 0.00 & 0.08 & -0.16 & -0.05 & -0.11 & 0.10 & - & - & - & - & - & - & \\    
UPK 219	& -290 & -0.54 & 1 & 0.05 & 0.05 & -0.20 & -0.09 & -0.23 & 0.00 & - & - & - & - & - & - & \\  
\hline
Stock 2 & -42 $\pm$ 238 & -0.23 $\pm$ 0.37 & 8 & 0.05 & 0.10 & -0.13 & 0.06 & -0.05 & 0.09 & 0.04 & -0.12 & -0.01 & -0.06 & 0.01 & - & \citet{Alonso2021} \\
\hline
\multicolumn{17}{c}{\Large \textbf{Literature}} \\
\hline
\multirow{2}{*}{Basel 11b} & - & - & - & 0.04 & -0.04 & -0.09 & 0.06 & 0.03  & 0.06  & -0.11 & -0.02 & 0.01  & - & - & - & \citet{Donor2020}  \\
& - & - & - & 0.07 & 0.02  & -0.07 & 0.06 & -0.06 & -0.01 & -0.14 & 0.02 & -0.02 & - & - & - & \citet{Spina2021} \\  
\hline
\multirow{3}{*}{Collinder 350} & 36 & 0.09 & 1 & -0.14 & -0.13 & -0.12 & 0.07 & - & -0.04 & - & -0.03 & 0.02  & - & 0.24 & 0.18 & \citet{Pakhomov2009} \\
& - & - & - & 0.07 & -0.16 & -0.15 & -0.07 & - & - & - & -0.08 & -0.07 & - & 0.11 & - & \citet{BlancoCuaresma2015} \\ 
& - & - & - & -0.06 & -0.01 & 0.02 & -0.03 & - & 0.10 & - & 0.01 & -0.03 & - & 0.11 & 0.18 & \citet{BlancoCuaresma2018}\\
\hline
Stock 2 & 36 $\pm$ 68 & 0.31 $\pm$ 0.13 & 2 & 0.04 & -0.13 & -0.17 & 0.07 & -0.18 & -0.10 & 0.04 & -0.10 & -0.02 & - & 0.22 & 0.01 & \citet{Reddy2019} \\
\hline
\multirow{2}{*}{NGC 2548} & 102 & -1.96 & 1 & - & - & - & - & - & - & - & - & - & - & - & - & \citet{Sun2020} \\  
& -24 & 0.17 & 1 & 0.04 & -0.01 & 0.00 & -0.03 & -0.11 & -0.11 & -0.02 & -0.06 & -0.05 & - & 0.26 & 0.11 & \citet{Spina2021} \\
\hline
\multirow{2}{*}{NGC 2632} & - & - & - & 0.03 & 0.07 & 0.07 & 0.08 & 0.20 & -0.05 & 0.06 & 0.08 & 0.12 & - & 0.08 & - & \citet{Spina2021} \\
& -144 $\pm$ 69 & -0.12 $\pm$ 0.17 & 2 & -0.07 & 0.01 & -0.03 & 0.08 & -0.06 & -0.18 & 0.16 & -0.07 & 0.05 & -0.19 & 0.05 & - & \citet{CarbajoHijarrubia2024} \\
\hline
\multirow{10}{*}{NGC 2682} & - & - & - & 0.07 & -0.07 & -0.07 & 0.10 & -0.13 & -0.13 & - & -0.06 & 0.05 & - & - & - & \citet{Tautvaiviene2000}\\
& - & - & - & 0.02 & -0.13 & -0.06 & 0.02 & -0.24 & -0.16 & - & - & 0.01 & - & - & - & \citet{Yong2006}\\
& - & - & - & -0.01 & -0.24 & -0.07 & 0.18 & -0.02 & -0.02 & - & -0.06 & 0.04 & - & 0.14 & - & \citet{Pancino2010} \\
& - & - & - & 0.01 & -0.02 & -0.15 & 0.28 & -0.07 & -0.10 & - & 0.18  & 0.11 & - & - & - & \citet{Friel2010} \\
& 22 $\pm$ 6 & 0.15 $\pm$ 0.09 & 2 & 0.05 & -0.20 & -0.18 & 0.24 & 0.03 & - & - & - & 0.10 & - & - & - & \citet{Jacobson2011} \\
& -87 $\pm$ 28 & 0.01 $\pm$ 0.08 & 2 & 0.08 & 0.03 & 0.01 & - & 0.03  & -0.00 & - & - & - & - & - & - & \citet{Gao2018} \\
& -32 $\pm$ 18 & 0.04 $\pm$ 0.02 & 2 & 0.01 & 0.02 & -0.02 & 0.10 & - & - & - & - & 0.03 & - & - & - &\citet{Casamiquela2017, Casamiquela2019} \\
& - & - & - & 0.04 & 0.03 & 0.02 & 0.14 & 0.01 & -0.03 & 0.02 & -0.28 & 0.10 & - & 0.06 & 0.04 & \citet{Spina2021} \\
& - & - & - & 0.03 & 0.03 & 0.03 & 0.15 & 0.02 & 0.02 & 0.02 & 0.04 & 0.07 & - & - & - & \citet{Donor2020} \\
& -94 $\pm$ 14 & 0.01 $\pm$ 0.01 & 2 & 0.01 & 0.02 & -0.08 & 0.10 & -0.15 & -0.14 & 0.11 & -0.03 & 0.02 & -0.12 & 0.16 & - & \citet{CarbajoHijarrubia2024} \\
\hline 
Ruprecht 171 & -220 $\pm$ 36 & -0.20 $\pm$ 0.08 & 3 & -0.22 & 0.01 & -0.02 & 0.02 & -0.10 & -0.13 & 0.03 & -0.02 & -0.07 & 0.07 & 0.15 & - & \citet{CarbajoHijarrubia2024} \\
\hline
Tombaugh 5 & 25 & 0.27 & 1 & -0.04 & - & -0.07 & -0.07 & - & - & - & - & -0.10 & - & - &  - & \citet{Baratella2018} \\ 
\hline
\end{tabular}
}
\small 
\\ \textbf{Notes:} For the \citet{Casali2020a}, the first difference value in $T_{\rm eff}$, $\log g$ and [Fe/H] is related to {\tt FAMA} and the second to {\tt ROTFIT} results.
\end{table*}

\begin{figure*}
\centering
\includegraphics[width=0.9\textwidth]{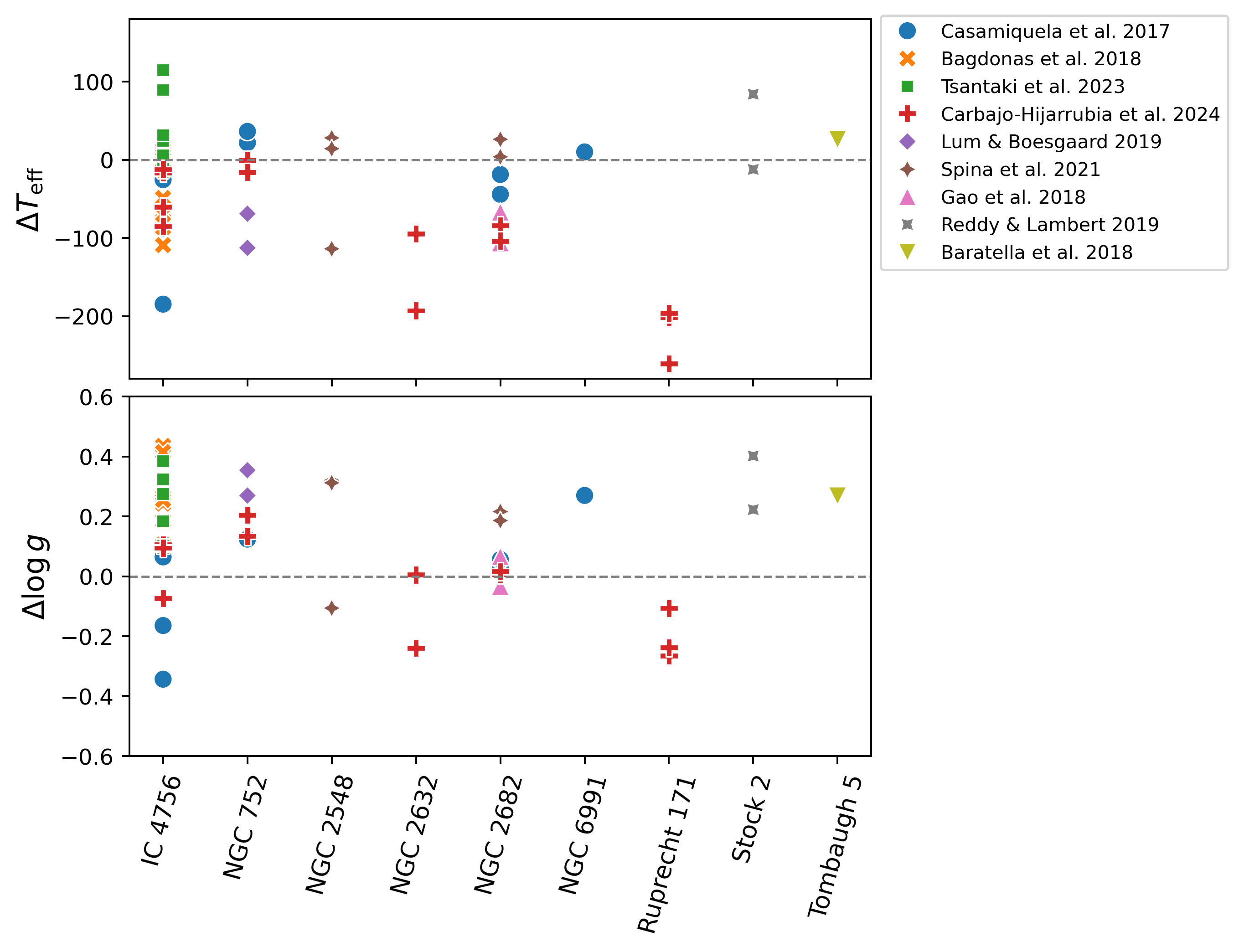}
\caption{Comparison of atmospheric parameters for stars matched with the literature, showing star-by-star differences.}
\label{fig:parameters_atm_stars}
\end{figure*}

\onecolumn
\newpage

\section{GALAH DR3 open clusters}

\begin{table}[ht]
\centering
\caption{Open clusters selected from \citet{Spina2021} with mean abundances exclusively from GALAH DR3.}
\label{tab:galah_oc}
\setstretch{1.2}
\scalebox{0.6}{
\begin{tabular}{lrrrrrrrrrrrr}
\hline \hline
\multicolumn{1}{c}{Cluster} & \multicolumn{1}{c}{[Fe/H]} & \multicolumn{1}{c}{[Mg/Fe]} & \multicolumn{1}{c}{[Si/Fe]} & \multicolumn{1}{c}{[Ti/Fe]} & \multicolumn{1}{c}{[Na/Fe]} & \multicolumn{1}{c}{[Al/Fe]} & \multicolumn{1}{c}{[Mn/Fe]} & \multicolumn{1}{c}{[Co/Fe]} & \multicolumn{1}{c}{[Ni/Fe]} & \multicolumn{1}{c}{[Sr/Fe]} & \multicolumn{1}{c}{[Y/Fe]} & \multicolumn{1}{c}{[Eu/Fe]} \\
\hline 
ASCC 16 & -0.369 $\pm$ 0.24 & -0.066 $\pm$ 0.10 & 0.161 $\pm$ 0.21 & -0.040 $\pm$ 0.04 & 0.165 $\pm$ 0.12 & 0.029 $\pm$ 0.01 & 0.089 $\pm$ 0.14 & - & -0.019 $\pm$ 0.01 & - & 0.341 $\pm$ 0.11 & - \\
ASCC 19 & 0.413 $\pm$ 0.41 & -0.049 $\pm$ 0.00 & -0.033 $\pm$ 0.01 & -0.122 $\pm$ 0.01 & -0.048 $\pm$ 0.01 & -0.054 $\pm$ 0.01 & 0.101 $\pm$ 0.00 & - & -0.213 $\pm$ 0.01 & - & 0.533 $\pm$ 0.01 & - \\
ASCC 99 & 0.152 $\pm$ 0.10 & -0.151 $\pm$ 0.15 & -0.081 $\pm$ 0.10 & -0.106 $\pm$ 0.14 & -0.089 $\pm$ 0.10 & -0.148 $\pm$ 0.11 & -0.060 $\pm$ 0.15 & 0.227 $\pm$ 0.21 & -0.060 $\pm$ 0.17 & - & -0.057 $\pm$ 0.23 & - \\
Alessi 9 & -0.065 $\pm$ 0.12 & -0.051 $\pm$ 0.07 & -0.076 $\pm$ 0.06 & 0.188 $\pm$ 0.19 & -0.164 $\pm$ 0.13 & 0.001 $\pm$ 0.01 & -0.073 $\pm$ 0.03 & -0.206 $\pm$ 0.01 & -0.141 $\pm$ 0.01 & - & 0.337 $\pm$ 0.14 & - \\
Alessi 24 & -0.137 $\pm$ 0.20 & -0.109 $\pm$ 0.06 & -0.030 $\pm$ 0.07 & 0.006 $\pm$ 0.04 & -0.005 $\pm$ 0.10 & -0.035 $\pm$ 0.07 & -0.059 $\pm$ 0.12 & 0.097 $\pm$ 0.14 & -0.042 $\pm$ 0.07 & - & 0.163 $\pm$ 0.24  & - \\
Berkeley 32 & -0.484 $\pm$ 0.14 & 0.050 $\pm$ 0.18 & 0.280 $\pm$ 0.20 & 0.176 $\pm$ 0.12 & 0.269 $\pm$ 0.21 & 0.200 $\pm$ 0.06 & 0.135 $\pm$ 0.19 & -0.026 $\pm$ 0.06 & 0.068 $\pm$ 0.15 & - & -0.018 $\pm$ 0.20 & 0.079 $\pm$ 0.08 \\
Berkeley 33 & -0.213 $\pm$ 0.10 & -0.037 $\pm$ 0.14 & -0.043 $\pm$ 0.09 & 0.291 $\pm$ 0.16 & 0.211 $\pm$ 0.09 & 0.070 $\pm$ 0.10 & -0.166 $\pm$ 0.14 & - & -0.081 $\pm$ 0.11 & 0.806 $\pm$ 0.20 & 0.401 $\pm$ 0.21 & -0.169 $\pm$ 0.12 \\
Berkeley 73 & -0.410 $\pm$ 0.14 & -0.070 $\pm$ 0.06 & 0.085 $\pm$ 0.04 & 0.437 $\pm$ 0.41 & 0.143 $\pm$ 0.08 & 0.314 $\pm$ 0.23 & -0.084 $\pm$ 0.06 & -0.102 $\pm$ 0.01 & 0.490 $\pm$ 0.33 & 0.866 $\pm$ 0.01 & 0.210 $\pm$ 0.01 & 0.589 $\pm$ 0.24 \\
BH 99 & -0.150 $\pm$ 0.01 & -0.072 $\pm$ 0.01 & 0.088 $\pm$ 0.03 & - & 0.052 $\pm$ 0.01 & -0.045 $\pm$ 0.04 & -0.007 $\pm$ 0.02 & - & - & - & 0.035 $\pm$ 0.05 & - \\
Blanco 1 & -0.054 $\pm$ 0.08 & -0.019 $\pm$ 0.10 & -0.038 $\pm$ 0.09 & 0.004 $\pm$ 0.12 & -0.114 $\pm$ 0.09 & -0.047 $\pm$ 0.07 & -0.033 $\pm$ 0.09 & 0.138 $\pm$ 0.20 & -0.068 $\pm$ 0.09 & 0.661 $\pm$ 0.01 & 0.198 $\pm$ 0.16 & - \\
Collinder 69 & -0.126 $\pm$ 0.22 & -0.075 $\pm$ 0.16 & 0.016 $\pm$ 0.13 & -0.031 $\pm$ 0.10 & 0.025 $\pm$ 0.17 & 0.072 $\pm$ 0.15 & 0.046 $\pm$ 0.11 & -0.264 $\pm$ 0.01 & -0.067 $\pm$ 0.20 & - & -0.033 $\pm$ 0.11 & 0.142 $\pm$ 0.01 \\
Collinder 135 & -0.032 $\pm$ 0.05 & -0.054 $\pm$ 0.08 & -0.056 $\pm$ 0.05 & 0.043 $\pm$ 0.07 & -0.075 $\pm$ 0.07 & -0.014 $\pm$ 0.05 & -0.007 $\pm$ 0.06 & 0.233 $\pm$ 0.01 & -0.136 $\pm$ 0.05 & - & 0.172 $\pm$ 0.14 & - \\
Collinder 261 & -0.048 $\pm$ 0.12 & -0.087 $\pm$ 0.19 & 0.025 $\pm$ 0.10 & -0.006 $\pm$ 0.18 & 0.034 $\pm$ 0.14 & 0.064 $\pm$ 0.13 & -0.086 $\pm$ 0.26 & -0.015 $\pm$ 0.16 & 0.011 $\pm$ 0.16 & 0.426 $\pm$ 0.23 & -0.023 $\pm$ 0.31 & 0.109 $\pm$ 0.16  \\
Collinder 359 & -0.077 $\pm$ 0.10 & -0.105 $\pm$ 0.07 & 0.096 $\pm$ 0.14 & -0.038 $\pm$ 0.01 & 0.164 $\pm$ 0.12 & 0.027 $\pm$ 0.10 & 0.135 $\pm$ 0.12 & - & -0.003 $\pm$ 0.01 & - & 0.094 $\pm$ 0.01 & - \\
Gulliver 6 & -0.006 $\pm$ 0.22 & -0.115 $\pm$ 0.12 & 0.089 $\pm$ 0.17 & 0.024 $\pm$ 0.10 & -0.048 $\pm$ 0.12 & -0.116 $\pm$ 0.11 & 0.042 $\pm$ 0.19 & - & -0.150 $\pm$ 0.04 & - & 0.165  $\pm$ 0.25 & - \\
Gulliver 13 & -0.249 $\pm$ 0.06 & 0.101 $\pm$ 0.01 & 0.008 $\pm$ 0.01 & 0.344 $\pm$ 0.01 & 0.140 $\pm$ 0.01 & 0.027 $\pm$ 0.01 & -0.233 $\pm$ 0.01 & - & - & - & 0.282  $\pm$ 0.01 & - \\
IC 2602 & -0.140 $\pm$ 0.17 & -0.014 $\pm$ 0.05 & -0.005 $\pm$ 0.05 & -0.007 $\pm$ 0.08 & 0.025 $\pm$ 0.12 & 0.006 $\pm$ 0.10 & 0.053 $\pm$ 0.05 & -0.210 $\pm$ 0.01 & 0.055 $\pm$ 0.11 & - & 0.179 $\pm$ 0.11 & - \\
IC 4665 & -0.134 $\pm$ 0.22 & -0.072 $\pm$ 0.10 & -0.019 $\pm$ 0.08 & -0.035 $\pm$ 0.07 & 0.048 $\pm$ 0.18 & -0.017 $\pm$ 0.06 & 0.083 $\pm$ 0.08 & - & -0.172 $\pm$ 0.05 & - & 0.281 $\pm$ 0.17  & -\\
LP 5 & -0.123 $\pm$ 0.01 & 0.098 $\pm$ 0.23 & 0.207 $\pm$ 0.07 & - & 0.232 $\pm$ 0.03 & 0.022 $\pm$ 0.01 & 0.351 $\pm$ 0.30 & -0.254 $\pm$ 0.01 & -0.011 $\pm$ 0.01 & - & -0.072 $\pm$ 0.01 & - \\
Mamajek 4 & 0.001 $\pm$ 0.14 & -0.102 $\pm$ 0.14 & 0.024 $\pm$ 0.09 & 0.069 $\pm$ 0.18 & 0.042 $\pm$ 0.13 & -0.011 $\pm$ 0.10 & 0.012 $\pm$ 0.09 & 0.154 $\pm$ 0.16 & -0.016 $\pm$ 0.10 & - & 0.148 $\pm$ 0.26 & - \\
Melotte 22 & -0.034 $\pm$ 0.15 & -0.033 $\pm$ 0.10 & -0.032 $\pm$ 0.09 & 0.030 $\pm$ 0.12 & -0.008 $\pm$ 0.13 & -0.033 $\pm$ 0.07 & 0.020 $\pm$ 0.08 & 0.139 $\pm$ 0.26 & -0.075 $\pm$ 0.22 & 0.182 $\pm$ 0.01 & 0.194 $\pm$ 0.12 & - \\
Melotte 25 & 0.120 $\pm$ 0.18 & -0.048 $\pm$ 0.18 & -0.119 $\pm$ 0.12 & 0.041 $\pm$ 0.14 & 0.048 $\pm$ 0.24 & 0.025 $\pm$ 0.08 & 0.049 $\pm$ 0.17 & -0.048 $\pm$ 0.13 & -0.022 $\pm$ 0.19 & - & 0.299 $\pm$ 0.33 & - \\
Melotte 101 & -0.154 $\pm$ 0.01 & 0.123 $\pm$ 0.01 & 0.057 $\pm$ 0.01 & -0.014 $\pm$ 0.01 & 0.350 $\pm$ 0.01 & 0.206 $\pm$ 0.01 & 0.057 $\pm$ 0.01 & -0.001 $\pm$ 0.01 & 0.125 $\pm$ 0.01 & - & 0.427 $\pm$ 0.01 & 0.013 $\pm$ 0.01 \\
NGC 1647 & 0.174 $\pm$ 0.01 & -0.209 $\pm$ 0.01 & -0.110 $\pm$ 0.01 & - & -0.205 $\pm$ 0.01 & - & - & - & - & - & - & -\\
NGC 1750 & -0.337 $\pm$ 0.14 & -0.182 $\pm$ 0.20 & 0.061 $\pm$ 0.16 & - & 0.170 $\pm$ 0.09 & - & 0.071 $\pm$ 0.04 & - & - & - & 0.500 $\pm$ 0.17 & - \\
NGC 1817 & -0.088 $\pm$ 0.11 & -0.068 $\pm$ 0.10 & 0.026 $\pm$ 0.12 & 0.011 $\pm$ 0.12 & 0.174 $\pm$ 0.11 & -0.007 $\pm$ 0.08 & -0.026 $\pm$ 0.12 & -0.106 $\pm$ 0.12 & 0.042 $\pm$ 0.12 & - & 0.122 $\pm$ 0.21 & 0.010 $\pm$ 0.13 \\
NGC 1901 & -0.272 $\pm$ 0.25 & -0.119 $\pm$ 0.14 & 0.014 $\pm$ 0.09 & 0.100 $\pm$ 0.12 & 0.066 $\pm$ 0.11 & -0.054 $\pm$ 0.10 & -0.045 $\pm$ 0.12 & - & -0.167 $\pm$ 0.01 & - & 0.197 $\pm$ 0.22 & - \\
NGC 2112 & -0.083 $\pm$ 0.08 & -0.000 $\pm$ 0.11 & -0.025 $\pm$ 0.09 & 0.045 $\pm$ 0.17 & -0.008 $\pm$ 0.12 & 0.023 $\pm$ 0.15 & -0.006 $\pm$ 0.12 & -0.062 $\pm$ 0.11 & -0.019 $\pm$ 0.09 & - & 0.010 $\pm$ 0.32 & 0.016 $\pm$ 0.10 \\
NGC 2204 & -0.325 $\pm$ 0.09 & -0.003 $\pm$ 0.07 & 0.043 $\pm$ 0.10 & 0.069 $\pm$ 0.11 & 0.242 $\pm$ 0.14 & 0.075 $\pm$ 0.12 & -0.019 $\pm$ 0.15 & -0.037 $\pm$ 0.09 & -0.007 $\pm$ 0.09 & 0.916 $\pm$ 0.01 & 0.227 $\pm$ 0.44 & 0.088 $\pm$ 0.08 \\
NGC 2215 & -0.100 $\pm$ 0.06 & 0.117 $\pm$ 0.06 & 0.022 $\pm$ 0.05 & -0.010 $\pm$ 0.05 & 0.010 $\pm$ 0.04 & -0.121 $\pm$ 0.06 & -0.009 $\pm$ 0.06 & - & -0.127 $\pm$ 0.07 & - & 0.004 $\pm$ 0.10 & - \\
NGC 2232 & -0.238 $\pm$ 0.19 & -0.048 $\pm$ 0.15 & 0.074 $\pm$ 0.06 & 0.040 $\pm$ 0.05 & 0.127 $\pm$ 0.11 & 0.069 $\pm$ 0.08 & 0.038 $\pm$ 0.09 & 0.775 $\pm$ 0.01 & -0.011 $\pm$ 0.11 & - & 0.254 $\pm$ 0.27 & - \\
NGC 2243 & -0.340 $\pm$ 0.12 & -0.027 $\pm$ 0.14 & 0.010 $\pm$ 0.11 & 0.117 $\pm$ 0.12 & 0.227 $\pm$ 0.07 & 0.022 $\pm$ 0.14 & 0.074 $\pm$ 0.18 & -0.088 $\pm$ 0.05 & 0.025 $\pm$ 0.13 & - & 0.039 $\pm$ 0.22 & 0.285 $\pm$ 0.06 \\
NGC 2318 & 0.002 $\pm$ 0.20 & - & - & - & 0.649 $\pm$ 0.10 & - & 0.134 $\pm$ 0.14 & - & - & - & - & -\\
NGC 2506 & -0.173 $\pm$ 0.12 & -0.071 $\pm$ 0.21 & -0.134 $\pm$ 0.12 & 0.214 $\pm$ 0.09 & 0.247 $\pm$ 0.12 & -0.007 $\pm$ 0.12 & 0.341 $\pm$ 0.20 & -0.087 $\pm$ 0.10 & 0.008 $\pm$ 0.14  & - & 0.288 $\pm$ 0.29 & 0.241 $\pm$ 0.14 \\
NGC 2516 & -0.137 $\pm$ 0.29 & -0.079 $\pm$ 0.12 & -0.004 $\pm$ 0.13 & -0.002 $\pm$ 0.12 & -0.082 $\pm$ 0.22 & -0.082 $\pm$ 0.09 & -0.009 $\pm$ 0.12 & 0.110 $\pm$ 0.25 & -0.101 $\pm$ 0.16 & - & 0.240 $\pm$ 0.25 & - \\
NGC 2539 & 0.023 $\pm$ 0.28 & -0.158 $\pm$ 0.09 & -0.101 $\pm$ 0.08 & -0.014 $\pm$ 0.08 & 0.095 $\pm$ 0.12 & -0.018 $\pm$ 0.06 & -0.107 $\pm$ 0.14 & -0.131 $\pm$ 0.14 & 0.009 $\pm$ 0.11 & - & 0.209 $\pm$ 0.23 & -0.040 $\pm$ 0.12 \\
NGC 2548 & -0.241 $\pm$ 0.26 & -0.045 $\pm$ 0.02 & -0.002 $\pm$ 0.02 & -0.018 $\pm$ 0.04 & 0.290 $\pm$ 0.11 & 0.053 $\pm$ 0.02 & 0.090 $\pm$ 0.28 & -0.068 $\pm$ 0.04 & 0.075 $\pm$ 0.04 & - & 0.546 $\pm$ 0.15 & -0.031 $\pm$ 0.03 \\
NGC 2632 & 0.198 $\pm$ 0.13 & -0.076 $\pm$ 0.09 & -0.057 $\pm$ 0.13 & -0.035 $\pm$ 0.08 & -0.006 $\pm$ 0.11 & -0.009 $\pm$ 0.08 & 0.051 $\pm$ 0.10 & 0.031 $\pm$ 0.22 & -0.010 $\pm$ 0.10 & 0.531 $\pm$ 0.18 & 0.090 $\pm$ 0.21 & - \\
NGC 2682 & -0.046 $\pm$ 0.16 & 0.016 $\pm$ 0.13 & 0.018 $\pm$ 0.14 & 0.055 $\pm$ 0.14 & 0.048 $\pm$ 0.12 & 0.030 $\pm$ 0.14 & 0.030 $\pm$ 0.16 & 0.319 $\pm$ 0.31 & 0.035 $\pm$ 0.12 & 0.548 $\pm$ 0.03 & 0.028 $\pm$ 0.21 & 0.013 $\pm$ 0.06 \\
NGC 3680 & -0.263 $\pm$ 0.07 & -0.283 $\pm$ 0.01 & 0.299 $\pm$ 0.01 & - & 0.172 $\pm$ 0.03 & - & 0.063 $\pm$ 0.01 & - & - & - & 0.329 $\pm$ 0.01 & - \\
NGC 5460 & -0.402 $\pm$ 0.01 & - & - & - & 0.273 $\pm$ 0.01 & - & - & - & - & - & -  & - \\
NGC 6253 & 0.159  $\pm$ 0.14 & 0.048 $\pm$ 0.10 & 0.115 $\pm$ 0.14 & 0.078  $\pm$ 0.18 & 0.308 $\pm$ 0.13 & 0.191 $\pm$ 0.19 & 0.205 $\pm$ 0.13 & 0.219 $\pm$ 0.10 & 0.159 $\pm$ 0.13 & 0.458 $\pm$ 0.32 & 0.015 $\pm$ 0.15 & -0.042 $\pm$ 0.05 \\
Ruprecht 145 & -0.257 $\pm$ 0.26 & -0.090 $\pm$ 0.06 & 0.058 $\pm$ 0.10 & -0.065 $\pm$ 0.04 & 0.256 $\pm$ 0.11 & 0.015 $\pm$ 0.05 & 0.053 $\pm$ 0.07 & -0.141 $\pm$ 0.04 & 0.015 $\pm$ 0.04 & - & 0.477 $\pm$ 0.34 & -0.153 $\pm$ 0.01 \\
Ruprecht 147 & 0.025 $\pm$ 0.14 & -0.038 $\pm$ 0.04 & -0.009 $\pm$ 0.07 & 0.002 $\pm$ 0.08 & 0.121 $\pm$ 0.07 & 0.051 $\pm$ 0.11 & 0.042 $\pm$ 0.10 & -0.037 $\pm$ 0.03 & 0.001 $\pm$ 0.10 & - & 0.083 $\pm$ 0.10 & -0.028 $\pm$ 0.05 \\
Trumpler 20 & -0.469 $\pm$ 0.17 & -0.304 $\pm$ 0.13 & 0.124 $\pm$ 0.13 & 0.021 $\pm$ 0.37 & 0.177 $\pm$ 0.13 & 0.443 $\pm$ 0.10 & -0.796 $\pm$ 0.11 & 0.442 $\pm$ 0.23 & 0.341 $\pm$ 0.17 & 0.733 $\pm$ 0.17 & -0.752 $\pm$ 0.11 & 0.225 $\pm$ 0.14 \\
Trumpler 26 & 0.190 $\pm$ 0.07 & -0.017 $\pm$ 0.08 & 0.002 $\pm$ 0.06 & -0.016 $\pm$ 0.05 & 0.464 $\pm$ 0.05 & 0.189 $\pm$ 0.07 & 0.181 $\pm$ 0.08 & -0.045 $\pm$ 0.06 & - & - & 0.374 $\pm$ 0.16 & -0.222 $\pm$ 0.10 \\
Turner 5 & -0.081 $\pm$ 0.07 & -0.055 $\pm$ 0.08 & 0.005 $\pm$ 0.06 & -0.008 $\pm$ 0.07 & -0.106 $\pm$ 0.06 & -0.038 $\pm$ 0.07 & -0.006 $\pm$ 0.08 & - & -0.062 $\pm$ 0.08 & - & -0.039 $\pm$ 0.14 & - \\
UBC 95 & -0.586 $\pm$ 0.30 & -0.052 $\pm$ 0.01 & 0.034 $\pm$ 0.01 & -0.357 $\pm$ 0.01 & 0.347 $\pm$ 0.01 & 0.404 $\pm$ 0.01 & -0.242 $\pm$ 0.01 & 0.248 $\pm$ 0.01 & 0.103 $\pm$ 0.01 & 0.460 $\pm$ 0.01 & -0.019 $\pm$ 0.01 & - \\
UBC 212 & -0.272 $\pm$ 0.09 & -0.139 $\pm$ 0.11 & 0.106 $\pm$ 0.08 & -0.040 $\pm$ 0.10 & 0.262 $\pm$ 0.07 & 0.103 $\pm$ 0.09 & -0.097 $\pm$ 0.11 & -0.049 $\pm$ 0.19 & -0.169 $\pm$ 0.10 & - & 0.293 $\pm$ 0.19 & 0.243 $\pm$ 0.11 \\
UBC 260 & -0.345 $\pm$ 0.04 & - & - & - & 0.352 $\pm$ 0.01 & - & - & - & - & - & 0.381 $\pm$ 0.01 & -  \\
UBC 511 & -0.415 $\pm$ 0.07 & -0.236 $\pm$ 0.05 & 0.147 $\pm$ 0.04 & - & 0.205 $\pm$ 0.03 & - & 0.483 $\pm$ 0.04 & - & - & - & 0.765 $\pm$ 0.07 & - \\
UPK 40 & 0.061 $\pm$ 0.09 & -0.064 $\pm$ 0.01 & 0.171 $\pm$ 0.01 & - & 0.104 $\pm$ 0.07 & - & 0.027 $\pm$ 0.01 & - & - & - & - & - \\
UPK 467 & -0.371 $\pm$ 0.49 & -0.027 $\pm$ 0.01 & 0.036 $\pm$ 0.01 & -0.008 $\pm$ 0.01 & 0.131 $\pm$ 0.01 & 0.225 $\pm$ 0.01 & 0.091 $\pm$ 0.01 & 0.039 $\pm$ 0.01 & 0.171 $\pm$ 0.01 & - & -0.129 $\pm$ 0.01 & - \\
UPK 524 & 0.008 $\pm$ 0.11 & 0.015 $\pm$ 0.04 & -0.015 $\pm$ 0.10 & 0.147 $\pm$ 0.01 & -0.045 $\pm$ 0.01 & -0.025 $\pm$ 0.14 & -0.012 $\pm$ 0.01 & - & -0.260 $\pm$ 0.01 & - & 0.214 $\pm$ 0.22 & - \\
UPK 526 & -0.147 $\pm$ 0.01 & -0.060 $\pm$ 0.01 & 0.026 $\pm$ 0.01 & 0.233 $\pm$ 0.01 & 0.056 $\pm$ 0.01 & 0.174 $\pm$ 0.19 & -0.055 $\pm$ 0.05 & - & - & - & 0.084 $\pm$ 0.07 & - \\
UPK 540 & 0.269 $\pm$ 0.15 & -0.481 $\pm$ 0.27 & - & - & -0.273 $\pm$ 0.15 & -0.466 $\pm$ 0.14 & -0.010 $\pm$ 0.25 & - & 0.275 $\pm$ 0.23 & - & - \\
UPK 545 & 0.260 $\pm$ 0.38 & -0.113 $\pm$ 0.17 & 0.144 $\pm$ 0.04 & 0.339 $\pm$ 0.01 & 0.110 $\pm$ 0.21 & 0.061 $\pm$ 0.21 & -0.009 $\pm$ 0.12 & -0.077 $\pm$ 0.01 & -0.557 $\pm$ 0.01 & - & 0.045 $\pm$ 0.01 & - \\
UPK 552 & -0.528 $\pm$ 0.12 & - & - & - & - &  - & - & - & - & - & 0.209 $\pm$ 0.19 & - \\
UPK 579 & -0.210 $\pm$ 0.07 & 0.009 $\pm$ 0.10 & 0.093 $\pm$ 0.07 & 0.120 $\pm$ 0.03 & 0.149 $\pm$ 0.03 & 0.075 $\pm$ 0.08 & -0.064 $\pm$ 0.10 & 0.815 $\pm$ 0.01 & 0.080 $\pm$ 0.01 & - & 0.278 $\pm$ 0.27 & - \\
UPK 587 & -0.557 $\pm$ 0.33 & -0.214 $\pm$ 0.14 & 0.169 $\pm$ 0.01 & - & 0.347 $\pm$ 0.16 & 0.376 $\pm$ 0.01 & 0.152 $\pm$ 0.04 & - & - & - & 0.421 $\pm$ 0.10 & - \\
UPK 599 & -0.375 $\pm$ 0.25 & -0.110 $\pm$ 0.01 & 0.016 $\pm$ 0.01 & 0.325 $\pm$ 0.01 & 0.046 $\pm$ 0.01 & -0.010 $\pm$ 0.01 & -0.090 $\pm$ 0.01 & - & - & - & 0.057  $\pm$ 0.01 & - \\
UPK 612 & -0.060 $\pm$ 0.21 & -0.097 $\pm$ 0.05 & -0.070 $\pm$ 0.04 & -0.025 $\pm$ 0.04 & -0.061 $\pm$ 0.03 & 0.009 $\pm$ 0.19 & -0.055 $\pm$ 0.08 & - & -0.115 $\pm$ 0.03 & - & 0.081  $\pm$ 0.05 & - \\
UPK 624 & -0.226 $\pm$ 0.02 & -0.143 $\pm$ 0.01 & 0.108 $\pm$ 0.01 & - & 0.142 $\pm$ 0.05 & - & -0.100 $\pm$ 0.05 & - & - & - & 0.141 $\pm$ 0.05 & -  \\
\hline
\end{tabular}
}
\end{table}

\end{appendix}

\end{document}